\documentclass{aa}  

\usepackage{graphicx}
\usepackage{tabularx} 

\usepackage{txfonts}
\usepackage{hyperref}
\usepackage{placeins}
\usepackage{txfonts}
\usepackage{colortbl}
\usepackage{orcidlink}
\usepackage{float}
\usepackage{lscape}
\usepackage{amsfonts,amsmath,natbib,xcolor,subfigure,gensymb,longtable,ulem}
\usepackage{hyperref}
\usepackage{siunitx}
\hypersetup{breaklinks,colorlinks,citecolor=blue,linkcolor=blue,urlcolor=red}
\usepackage[capitalise]{cleveref}
\defcitealias{Camisasca23}{C23}
\defcitealias{Golkhou15}{G15}
\defcitealias{Veres23}{V23}

\newcommand{\redcross}{\textcolor{red}{\(\times\)}}
\newcommand{\greencheck}{\textcolor{green}{\(\checkmark\)}}

\begin{document}

   \title{GRB minimum variability timescales with \textit{Fermi}/GBM}

   \subtitle{}

\author{R.~Maccary~\inst{1,2}\fnmsep\thanks{\texttt{mccrnl[at]unife[dot]it}} \orcidlink{0000-0002-8799-2510}\and   
            C.~Guidorzi\inst{1,2,3}\orcidlink{0000-0001-6869-0835}\and
           A.~E.~Camisasca~\inst{8,1}\orcidlink{0000-0002-4200-1947}\and
           M.~Maistrello~\inst{1,2}\orcidlink{0009-0000-4422-4151}\and
            S.Kobayashi~\inst{7}\orcidlink{0009-0000-4422-4151} 
           L.~Amati~\inst{2}\orcidlink{0000-0001-5355-7388}\and 
            L.~Bazzanini~\inst{1,2}\orcidlink{0000-0003-0727-0137}\and
           M.~Bulla~\inst{1,3,4}\orcidlink{0000-0002-8255-5127}\and
           L.~Ferro~\inst{1,2}\orcidlink{0009-0006-1140-6913}\and
           F.~Frontera~\inst{1,2}\orcidlink{0000-0003-2284-571X}\and
           A.~Tsvetkova~\inst{5,6}\orcidlink{0000-0003-0292-6221}
          }
          
   \institute{
   Department of Physics and Earth Science, University of Ferrara, via Saragat 1, I--44122, Ferrara, Italy\label{unife}\and 
   INAF -- Osservatorio di Astrofisica e Scienza dello Spazio di Bologna, Via Piero Gobetti 101, I-40129 Bologna, Italy\label{oabo}\and 
   INFN -- Sezione di Ferrara, via Saragat 1, I--44122, Ferrara, Italy\label{infnfe}\and
   INAF, Osservatorio Astronomico d’Abruzzo, via Mentore Maggini snc, 64100 Teramo, Italy\and
   Department of Physics, University of Cagliari, SP Monserrato-Sestu, km 0.7, 09042 Monserrato, Italy\and
   Ioffe Institute, Politekhnicheskaya 26, 194021 St. Petersburg, Russia\and
    Astrophysics Research Institute, Liverpool John Moores University, Liverpool Science Park IC2, 146 Brownlow Hill, Liverpool L3 5RF, UK\and Alma Mater Studiorum, Università di Bologna, Dipartimento di Fisica e Astronomia (DIFA), Via Gobetti 93/2, 40129 Bologna, Italy\\
   }
   
 
  \abstract
   {Gamma-ray bursts (GRBs) have traditionally been classified by duration into long (LGRBs) and short (SGRBs), with the former believed to originate from massive star collapses and the latter from compact binary mergers. However, events such as the SGRB 200826A (coming from a collapsar) and the LGRBs 211211A and 230307A (associated with a merger) suggest that duration-based classification could be sometimes misleading.
   Recently, the minimum variability timescale (MVT) has emerged as a key metric for classifying GRBs.
   }
   {We calculate the MVT, defined as the full width at half maximum (FWHM) of the narrowest pulse in the light curve, 
   using an independent dataset from \textit{Fermi}/GBM and we compare our results with other MVT definitions. We update the MVT-$T_{\rm{90}}$ plane and analyse peculiar events like long-duration merger candidates 211211A, 230307A, and other short GRBs with extended emission (SEE-GRBs). We also examine extragalactic magnetar giant flares (MGFs) and explore possible new correlations with peak energy.
   }
   {We used the {\sc mepsa} algorithm to identify the shortest pulse in each GRB light curve and measure its FWHM. We calculated the MVT for around 3700 GRBs, 177 of which with spectroscopically known redshift.
   }
   {SEE-GRBs and SGRBs share similar MVTs (from few tens to a few hundreds of ms), indicating a common progenitor, while extragalactic MGFs exhibit even shorter values (from few ms to few tens of ms). Our MVT estimation method consistently yields higher values than another existing technique, the latter aligning with the pulse rise time. For LGRBs, we confirmed the correlations of MVT with peak luminosity and Lorentz factor.
   }
   {We confirmed that, although MVT alone cannot determine the GRB progenitor, it is a valuable tool when combined with other indicators, helping to flag long-duration mergers and distinguish MGFs from typical SGRBs.}


   \maketitle
%
\section{Introduction}
\label{sec:intro}

Gamma-ray bursts (GRBs) are brief yet extremely intense flashes of gamma rays produced at cosmological distances. They are thought to arise from at least two types of catastrophic events: (i) the core-collapse of certain types of massive star, known as collapsars \citep{Woosley93,Paczynski98,MacFadyen99,Yoon05} - typically occurring at the centre of star-forming galaxies \citep{Fruchter06}, and associated with Type Ic-BL supernovae \citep{Galama98,Hjorth03}, and (ii) binary compact object mergers \citep{Blinnikov1984,Paczynski86, Eichler89,Paczynski91,Narayan92,LIGO-Fermi17}. The central engine, which powers an ultra-relativistic jet, could either be a stellar-mass black hole, surrounded by a hyper-accreting thick accretion disk \citep{Popham99,DiMatteo02,Janiuk07,Lei13}, or a strongly magnetised, rapidly spinning neutron star, also known as a magnetar \citep{Usov92,Wheeler00,Thompson04,Metzger11}. Despite much progress, the exact nature of the central engine(s), the mechanism(s) by which the relativistic jet is launched, the jet composition, and the radiation process(es) responsible for the gamma-ray emission remain unresolved questions.

Initially, GRBs were classified by their duration, with long GRBs (LGRBs) typically linked to collapsars and short GRBs (SGRBs) to mergers. 
However, a class of events known as SGRBs with extended emission (SEE-GRBs), presenting a short, hard spike followed by a longer, softer emission (sometimes lasting tens of seconds) challenged this simple classification \citep{Norris06}. For instance, events like 060614 \citep{Gehrels06,DellaValle06,Fynbo06,Jin15,Yang15}, which exhibited zero spectral lag and no evidence of a supernova despite occurring at low redshift, raised doubts about the reliability of using duration alone to infer the kind of progenitor. In many cases, the host galaxy remains undetected, and redshift measurements are unavailable, making it difficult to determine the progenitor type.
Recent cases, such as 211211A ($T_{90} \simeq 34~{\rm s}$;    \citealt{Rastinejad22,Gompertz23,Yang22,Troja22,Xiao22}) and 230307A ($T_{90} \simeq 35~{\rm s}$; \citealt{Dalessi23,Xiong23,Du24a,Dai24,Levan24,Yang24}) were followed by a kilonova, which provided compelling evidence that even mergers can produce long-duration GRBs, further emphasising the need for a new classification system. Given the growing complexity, families (i) and (ii) are now frequently described as merger GRBs and collapsar GRBs, or alternatively, as Type-I and Type-II GRBs, respectively \citep{Zhang06_nat}.

Among fast high-energy transient events, there is another category, known as magnetar giant flares (MGFs), which is sometimes mistaken for typical SGRBs. These events, produced by galactic or extragalactic magnetars, are characterised by a shorter rise time and duration, a harder peak energy, and a lower equivalent-isotropic energy $E_{\rm iso} \approx 10^{44-46}~\rm{erg}$ compared to cosmological SGRBs. When occurring within the Milky Way, they exhibit a long decaying tail, modulated by the neutron star rotation period \citep{Mazets79,Feroci99,Hurley99,Hurley05}, which is below instrumental sensitivity when they happen in nearby galaxies \citep{Ofek06,Mazets08,Burns21,Svinkin21,Roberts21,LAT21,Mereghetti24,Trigg24,Rodi25}. 

Numerous attempts to classify GRBs using different prompt emission properties have been made \citep{Goldstein10,Lu10,Lu14b,Tsvetkova25}. Many efforts have been made to develop machine-learning (ML) based GRB classification methods \citep{Jespersen20,Salmon22,Steinhardt23,Dimple23,Garcia-Cifuentes23,Chen24,Zhu24,Dimple24}. These methods generally recover the usual properties of the two GRB classes: ML-identified Type I GRBs tend to be shorter and spectrally harder than ML-identified Type II GRBs. However, complex cases—such as GRB 211211A and GRB 230307A—continue to challenge even the most advanced classification algorithms (see e.g. \citealt{Zhu24}). This highlights the persistent challenges in GRB classification, and the importance of identifying the most relevant parameters for distinguishing GRB progenitors.
A promising approach involves the minimum variability timescale (MVT), defined as the shortest timescale over which the signal shows uncorrelated temporal variability. Several methods have been proposed to calculate the MVT, such as temporal deconvolution into pulses \citep{Norris96,Norris05,Bhat12}, or wavelet decomposition \citep{MacLachlan13,Golkhou14,Golkhou15,Vianello18}.

The MVT could be directly linked to the activity of the central engine, as is the case for the internal shocks model (IS, \citealt{Rees94, Kobayashi97,Daigne98}), or may originate locally in the emission region. In the latter case, either relativistic turbulence \citep{Kumar09}, or the emission of Doppler-boosted local emitters \citep{Lyutikov03} could determine the MVT. These two pictures are being unified by the Internal Collision-Induced Magnetic Reconnection and Turbulence model (ICMART; \citealt{ICMART}), according to which longer timescales are linked to the central engine activity, while the shorter ones are attributed to relativistic magnetic turbulence within the emission region.

\citet[hereafter C23]{Camisasca23} defined the MVT as the full width half maximum (FWHM) of the shortest pulse that is detected with statistical confidence within a GRB light curve (LC).
This method builds on the {\sc mepsa} algorithm \citep{Guidorzi15a}, designed to identify statistically significant peaks in a given GRB LC. This method has the advantage of having a straightforward interpretation. In their study, they explored various possible correlations between the MVT, Lorentz factor, jet opening angle, and peak luminosity.
 
This method was also applied to the case of 230307A, where a MVT of 28 ms was reported by \citet{Camisasca23b}, suggesting a merger origin, in agreement with the discovery of a kilonova \citep{Bulla23b,Levan24}. This confirms the usefulness of the MVT in identifying long-duration merger candidates. The MVT can also be useful in distinguishing extragalactic MGFs from regular SGRBs.

The combination of MVT and other metrics may help further identify interesting merger candidates: in fact, \citet{Guidorzi24b} showed that the combination of high variability ($V > 0.1$), relatively low luminosity $L_{\rm iso} < 10^{51}~\rm{erg~s^{-1}}$ and short MVT ($\leq 0.1~\rm{s}$) may be indicative of a compact binary merger origin, in spite of the long duration and misleading temporal profile.

Our goal is to verify the results obtained by \citetalias{Camisasca23} using the complementary dataset of \textit{Fermi} Gamma-ray Burst Monitor (GBM; \citealt{Meegan09}). This is an all-sky monitor and is sensitive to soft gamma-rays, with 12 NaI scintillators working in the range from 8 to 1000 keV, with two additional BGO detectors operating from 150 keV to 30 MeV .

On the one hand it is important to test the results obtained by \citetalias{Camisasca23} through an independent data set. On the other hand, the GBM data in particular allow us to update and extend the analysis to interesting candidates that were detected exclusively with GBM. With over 3000 recorded GRBs, excellent time resolution ($<10~\rm{\mu s}$), and its large energy passband, \textit{Fermi}/GBM is ideally suited to conduct a statistical analysis of GRB MVTs.
Section \ref{sec:data_an} describes the GRB sample and the data analysis. Results are reported in  Section~\ref{sec:Results}. We
discuss the implications and conclude in Section~\ref{sec:discussions.}.
Hereafter, we used the flat-$\Lambda$CDM cosmology model with the latest cosmological parameters values $H_0 =67.66~\rm{km~Mpc^{-1}~s^{-1}}$ and 
$\Omega_0 = 0.31$ \citep{cosmoPlanck20}.
\section{Data analysis}
\label{sec:data_an}
\subsection{Dataset}
\label{sec:data_set}
We started with 3792 GRBs triggered
by \textit{Fermi}/GBM from 14 July 2008 to 11 June 2024, keeping 3720 of them, the others being not entirely covered by time tagged events (TTE) data. Some very bright GRBs, such as 221009A and 130427A were also removed owing to their brightness, which saturated the NaI detectors \citep{Ackermann14,Lesage23}. Among the remaining GRBs, 177 have measured redshift, with 152 classified as collapsar-candidate (or Type-II), 20 as merger candidates (or Type-I), and 5 as SEE-GRBs. 17 GRBs of the former class are associated with a supernova (SN). We also have 44 SEE-GRBs, either identified by \citet{Kaneko15,Lien16,Lan20}, or reported as such by the Gamma-Ray Burst Coordinate Network (GCN). Additionally, two long-duration merger candidates 211211A and 230307A appear in our sample. Those are LGRBs, having $T_{\rm{90}}>2~\rm{s}$, and not necessarily following the morphology of SEE-GRBs.
Table~\ref{tab:dataset} reports the data.

\begin{table}[h!]
\centering
\renewcommand{\arraystretch}{1.5} 
\resizebox{0.5\textwidth}{!}{%
\begin{tabular}{c c c c c c c}
\hline
\hline
GRB & Fermi Id & FWHM$_{\text{min}}$ [s] & T$_{90}$ [s] & $z$ & N$_{\text{p}}$ & Type\\
\hline
080714B & bn080714086 & $2.429^{+0.848}_{-0.628}$ & $5.376$ & - & $2$ & II \\
080714C & bn080714425 & $5.387_{-1.393}^{ 1.880}$ &  $40.192$ & -& 2& II \\
080714A  & bn080714745 & $3.825_{-0.990}^{1.335}$ & $59.649$ & - & 1& II \\
080715 & bn080715950 & $0.172_{-0.045}^{0.060}$ &	$7.872$ & - &2 & II \\
080717 & bn080717543 &	$7.252_{-1.876}^{2.532}$ & $36.608$ & - &1 & II \\
\hline
\\
\end{tabular}
}
\caption{Five first GRBs of our sample. This table is available in its entirety on machine-readable form (see Section~\ref{sec:data_av}.)}
\label{tab:dataset}
\end{table}

We also considered three extragalactic magnetar giant flares (MGF):  200415A \citep{Yang20,Roberts21,Svinkin21,LAT21}, 231115A \citep{Mereghetti24,MinaevPozanenko24}, and 180128A \citep{Trigg24}. Data are reported in Table~\ref{tab:data_MGF}. 
\begin{table}[h!]
\centering
\renewcommand{\arraystretch}{1.5}  
\resizebox{0.5\textwidth}{!}{%
\begin{tabular}{c c c c c c}
\hline
\hline
GRB & Fermi Id & FWHM$_{\text{min}}$ [ms] & T$_{90}$ [s] & $d$ [Mpc]  \\
\hline
180128A  & bn180128215  & $ 8.12_{-2.1}^{+2.80}$ & $0.208$ & 3.7   \\
200415A & bn200415367 & $ 2.97 _{ -0.77}^{+1.04}$ &  $0.144$ & 3.5  \\
231115A & bn231115650 & $24.41^{+8.52}_{ -6.31 }$ & $0.032$ & 3.5   \\
\hline
\\
\end{tabular}
}
\caption{Three MGFs of our sample.}
\label{tab:data_MGF}
\end{table}

\subsection{Data reduction}
\label{sec:bkgd}
We used the TTE data in the 8-1000 keV energy range, with an integration time of 64~ms. Whenever it was required by the procedure described in \citetalias{Camisasca23}, we also used 1024, 4, and 1~ms. For each burst, we selected the GBM detectors based on the ``bcat detector mask'' entry in the HEASARC  catalogue\footnote{\url{https://heasarc.gsfc.nasa.gov/db-perl/W3Browse/w3table.pl?tablehead=name\%3Dfermigbrst&Action=More+Options}}.
 We discarded the GRBs affected by solar flares and those with profiles not entirely contained in the TTE mode of GBM.
Background was interpolated and subtracted using the GBM data tools\footnote{\url{https://fermi.gsfc.nasa.gov/ssc/data/analysis/gbm/gbm_data_tools/gdt-docs/}.}
\citep{GbmDataTools}, following standard procedures also applied in \citet{Maccary24}.
\subsection{Minimum variability timescale computation}
\label{sec:mvt}
We adopted the MVT calculation defined in \citetalias{Camisasca23} as the FWHM of the narrowest, statistically significant peak in the LC (denoted hereafter as $\rm{FWHM}_{\rm{min}}$).
We measured $\rm{FWHM}_{\rm{min}}$ following the prescriptions of \citetalias{Camisasca23}, which build upon {\sc mepsa}:
\begin{itemize}
    \item the MVT is tentatively computed on the 64 ms LC, and the binning scheme is refined down to 4 ms or even 1 ms when needed;
    \item {\sc mepsa} is applied to the corresponding LCs, using a maximum rebin factor of 256;
    \item Peaks are filtered, using S/N thresholds following a scheme ensuring the same false alarm probability through different bin times\footnote{S/N $\geq$ 7, 6.8, 6.4, 6 at 1, 4, 64, 1024 ms} (see Figure 1 of \citetalias{Camisasca23}).
    \item The FWHM of each peak is computed using the calibrated formula established in \citetalias{Camisasca23}{}, which depends on {\sc mepsa} parameters. Then, the FWHM of the shortest significant peak is defined as ${\rm FWHM_{min}}$.
\end{itemize}
We compared the results of this method with a more direct computation of the FWHM of the shortest significant pulse obtained by fitting its time profile with a fast rise exponential decay (FRED) model (see Appendix~\ref{sec:appendix} for more details). A comparison of the results obtained using \textit{Swift}/BAT and \textit{Fermi}/GBM data was also performed in Appendix~\ref{sec:appendix_B}. We also compared in Appendix~\ref{sec:appendix_bb} the results of our method with those obtained with the Bayesian Blocks (BBs; \citet{Scargle13}) algorithm.

Out of 3720 GRBs, we obtained 3350 GRBs with a reliable measure of $\rm{FWHM}_{\rm{min}}$; 2992 of them are LGRBs ($T_{90}>2~\rm{s}$), while 358 are SGRBs ($T_{90}<2~\rm{s}$). For 29 GRBs, we could only determine an upper limit to $\rm{FWHM}_{\rm{min}}$. For 339 GRBs, the signal-to-noise ratio (S/N) was not high enough to enable a reliable measure of $\rm{FWHM}_{\rm{min}}$.
\section{Results}
\label{sec:Results}
In the following, we present the results of the MVT measurements for the 3350 GRBs for which $\rm{FWHM}_{\rm{min}}$ was successfully determined. We analyse the distribution of MVT of different GRB classes, examine correlations with other burst properties, and compare these findings with models and former studies.

\subsection{Comparison between our MVT estimation with other techniques}
\label{sec:comparison}
In this work, we compared three possible ways to measure the MVT: 
\begin{itemize}
    \item (i) the FWHM of the narrowest pulse, denoted as $\rm{FWHM_{min}}$;
    \item (ii) the MVT as computed by \citet[hereafter G15]{Golkhou15} and \citet[hereafter V23]{Veres23} using wavelets decomposition, denoted as $\Delta t_{\rm{min}}$; \item (iii) the detection timescale of the narrowest pulse identified by {\sc mepsa}, denoted as $\Delta t_{\rm det}$; this is the time interval over which the detection significance is maximised (see \citealt{Guidorzi15a} for details).
\end{itemize}

$\rm{FWHM_{min}}$ can be computed either by directly fitting the LC, or by using {\sc MEPSA} along with the procedure described in Sec.~\ref{sec:mvt}. As demonstrated in Appendix \ref{sec:appendix}, both methods give similar results, with most estimates compatible within uncertainties. Therefore, for the purpose of (i), we will only consider the latter, as it gives a reliable estimate of the narrowest pulse FWHM.

\begin{figure}[h!]
    \centering
\includegraphics[width=0.43\textwidth]{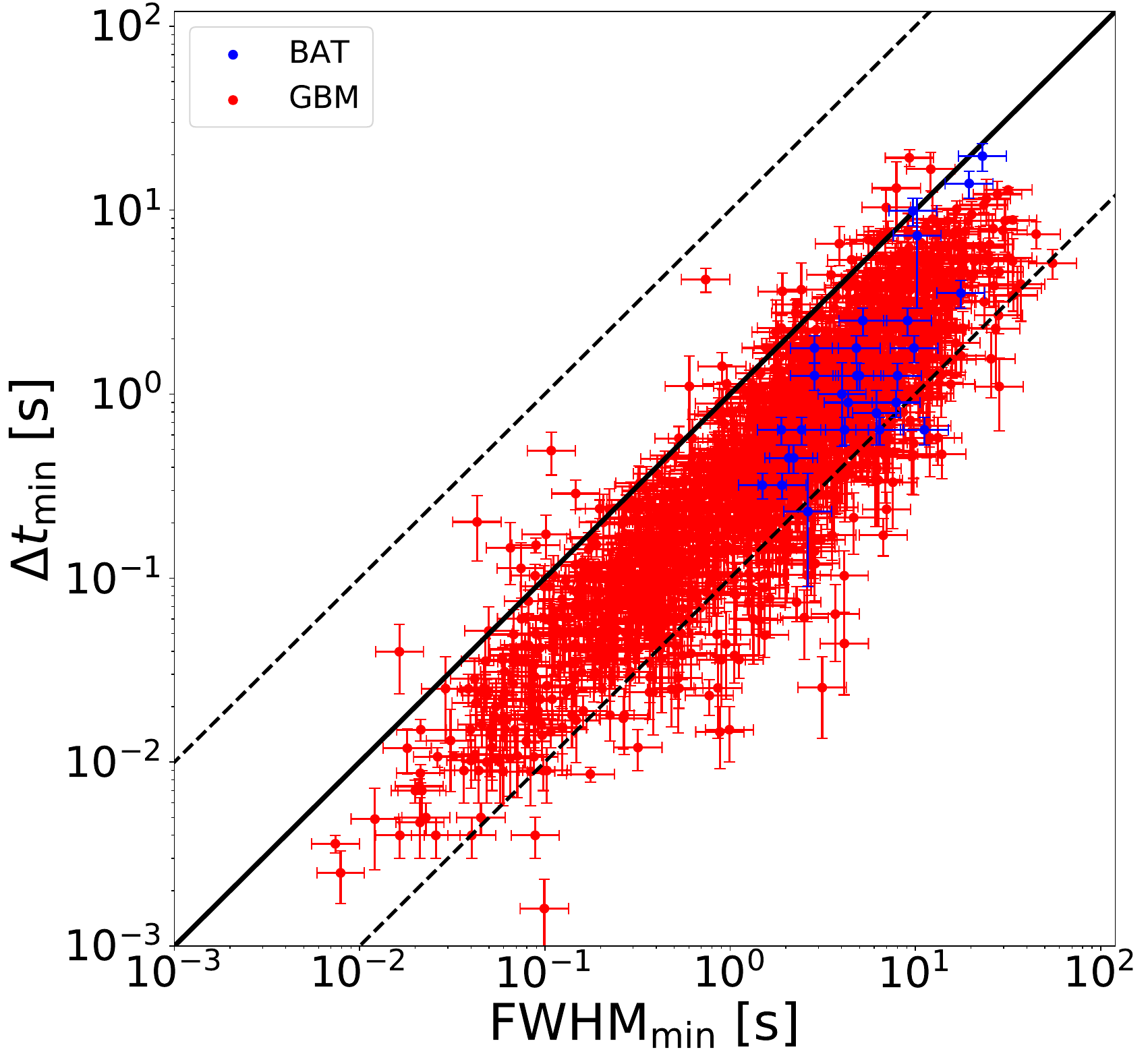}
    \caption{ This plot represents $\Delta t_{\rm{min}}$ versus $\rm{FWHM_{min}}$, for the GRBs in common. Red points show GBM data, where $\Delta t_{\rm{min}}$ was taken from \citetalias{Golkhou15} and \citetalias{Veres23}, while blue points are BAT data, $\Delta t_{\rm{min}}$ being taken from \citet{Golkhou14}. Equality is shown with a solid line, while dashed lines show $\pm 1$~dex.}
\label{fig:FWHM_min_vs_dt_min}
\end{figure}

\begin{figure}[h!]
    \centering
\includegraphics[scale=0.24]{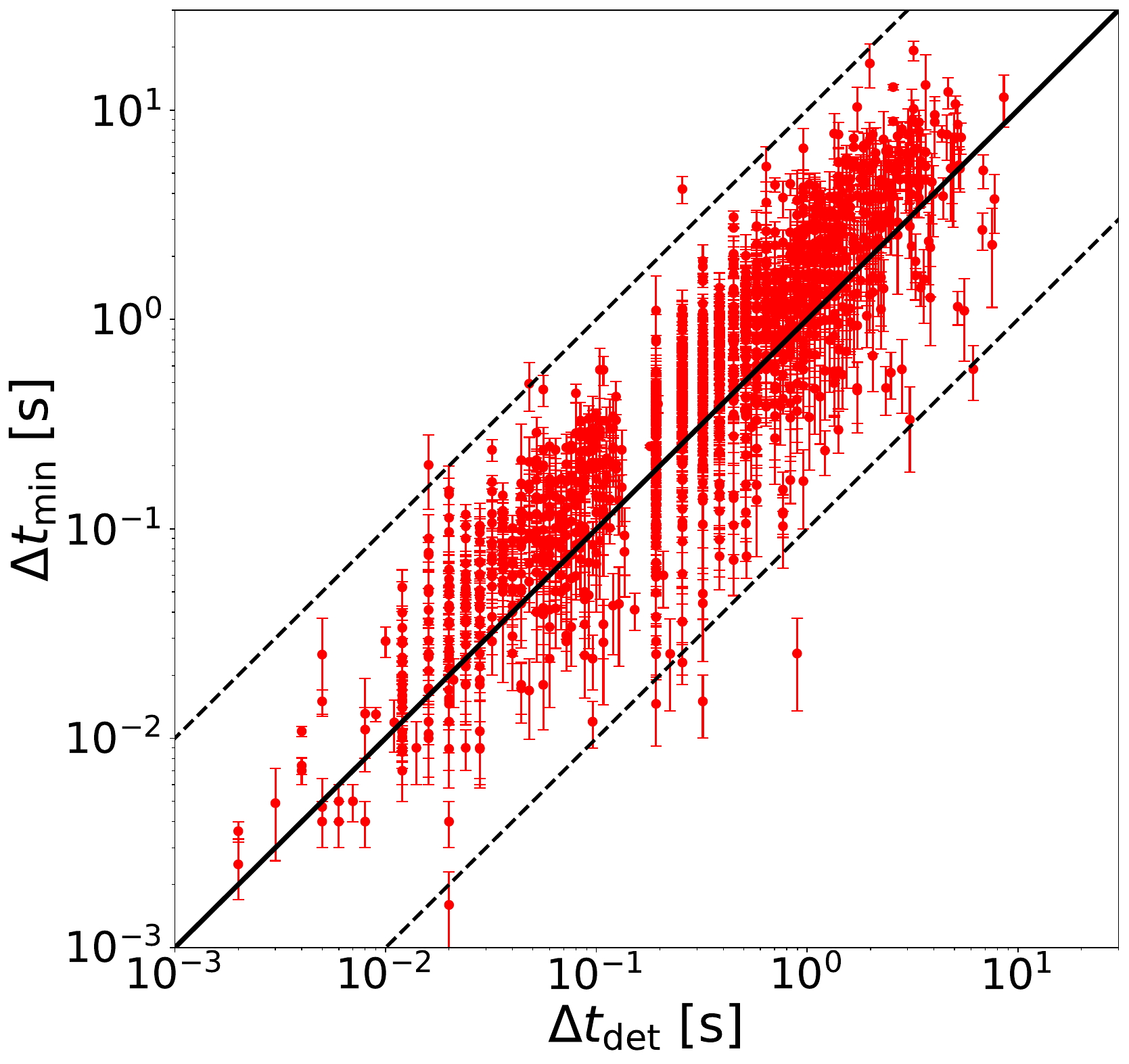}
    \caption{$\Delta t_{\rm min}$ is the MVT estimate from \citetalias{Golkhou15} and \citetalias{Veres23} obtained with GBM, while $\Delta t_{\rm det}$ is the detection timescale found with {\sc mepsa}. Solid and dashed lines have the same meaning as in Figure ~\ref{fig:FWHM_min_vs_dt_min}.}
\label{fig:det_timescale}
\end{figure}

Figure~\ref{fig:FWHM_min_vs_dt_min} compares $\rm{FWHM_{min}}$ and $\Delta t_{\rm min}$. We observe a significant discrepancy between the two, $\rm{FWHM_{min}}$ being significantly longer. Interestingly, $\Delta t_{\rm min}$ values are consistent with the detection timescale values $\Delta t_{\rm det}$ calculated by {\sc mepsa} (Figure ~\ref{fig:det_timescale}).

According to simulations carried out in \citet{Guidorzi15a} and \citetalias{Camisasca23}, the brighter the pulse, the smaller the ratio $\Delta t_{\rm det}$/FWHM. Specifically, \citetalias{Camisasca23} came up with a calibrated relation between FWHM$_{\rm min}$ and $\Delta t_{\rm det}$, which they used to estimate the former from the latter, also using other ancillary information yielded by {\sc mepsa} and modelling the corresponding uncertainty (see equation~A3 therein).

\begin{figure}
    \centering    \includegraphics[width=0.44\textwidth]{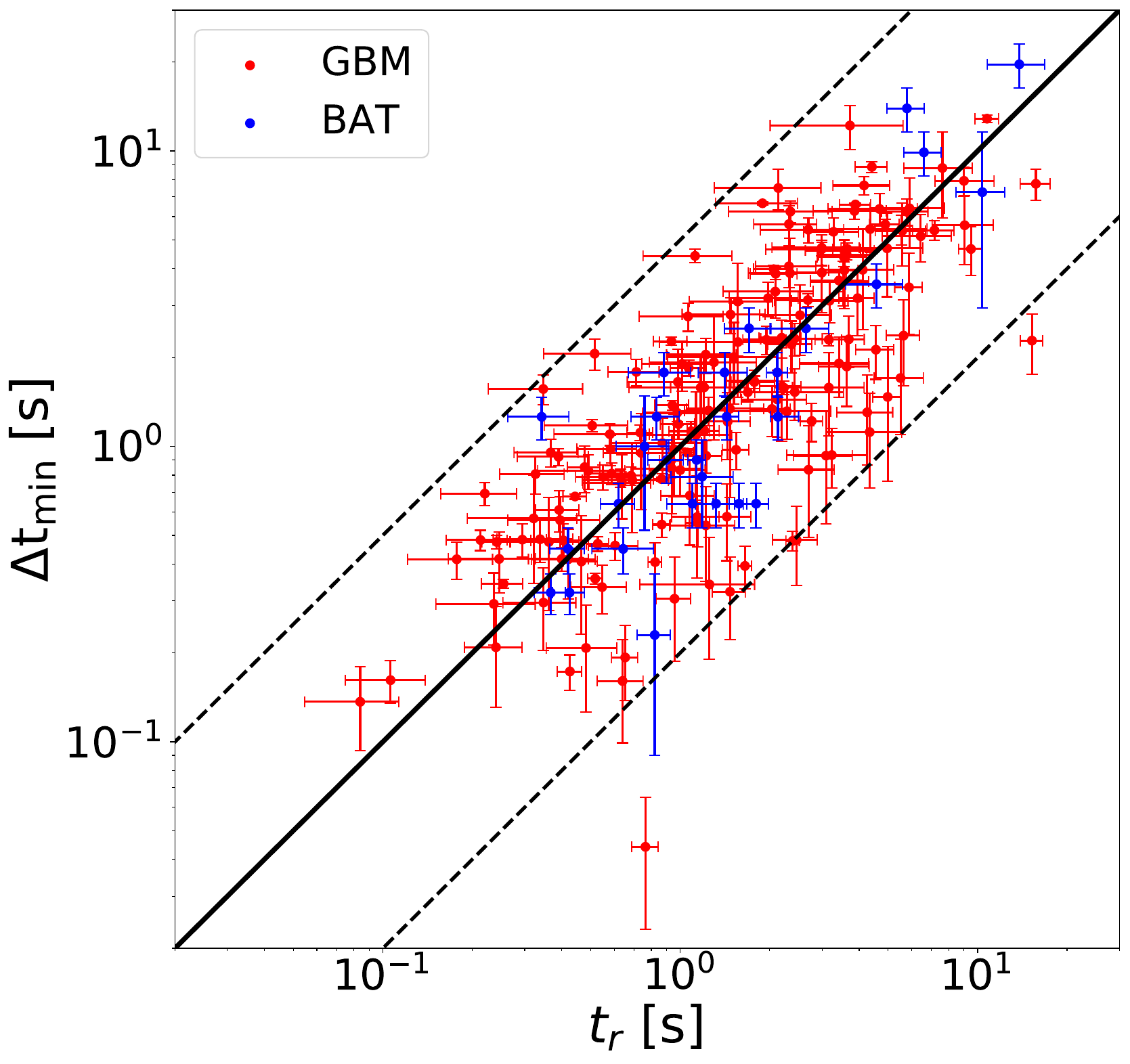}
    \caption{$\Delta t_{\rm{min}}$ vs. the rise time $t_r$ of the fitted pulse, for the samples of GRBs defined in Appendix \ref{sec:appendix}. Red points show GBM data, where $\Delta t_{\rm{min}}$ was taken from \citetalias{Golkhou15} and \citetalias{Veres23}, while blue points are BAT data, $\Delta t_{\rm{min}}$ being taken from \citet{Golkhou14}.  Solid and dashed lines have the same meaning as in Figure ~\ref{fig:FWHM_min_vs_dt_min}.}
    \label{fig:tr_vs_FWHMmin}
\end{figure}

Our analysis confirmed that $\Delta t _{\rm{min}}$ is more closely related with the detection timescale (Figure ~\ref{fig:det_timescale}) and the rise time $t_r$ (Figure ~\ref{fig:tr_vs_FWHMmin}), rather than the FWHM (Figure ~\ref{fig:FWHM_min_vs_dt_min}) of the pulse, as also mentioned by \citetalias{Golkhou15}. Indeed, $\Delta t_{\rm{det}} \sim \Delta t_{\rm{min}} \sim t_r $, with median values of $0.6$ and $1.1$ for $\Delta t_{\rm{det}}/\Delta t_{\rm{min}}$ and $\Delta t_{\rm{min}}/t_r$, respectively.
GRB pulses are typically asymmetric, with a decay-to-rise time ratio of 3-4: as a consequence, the FWHM is comparably longer than the rise time alone and explains why $\rm{FWHM_{min}}$ is longer than $\Delta t_{\rm min}$ by a comparable factor.

This difference between $\rm{FWHM_{min}}$ and $\Delta t_{\rm min}$ is important to bear in mind, especially when it is to be interpreted within a theoretical context.

\subsection{$\rm{FWHM}_{\rm{min}}$-$T_{\rm{90}}$ plane}
\label{sec:FWHM_vs_T90}
\begin{figure*}
    \centering
\includegraphics[scale=0.7]{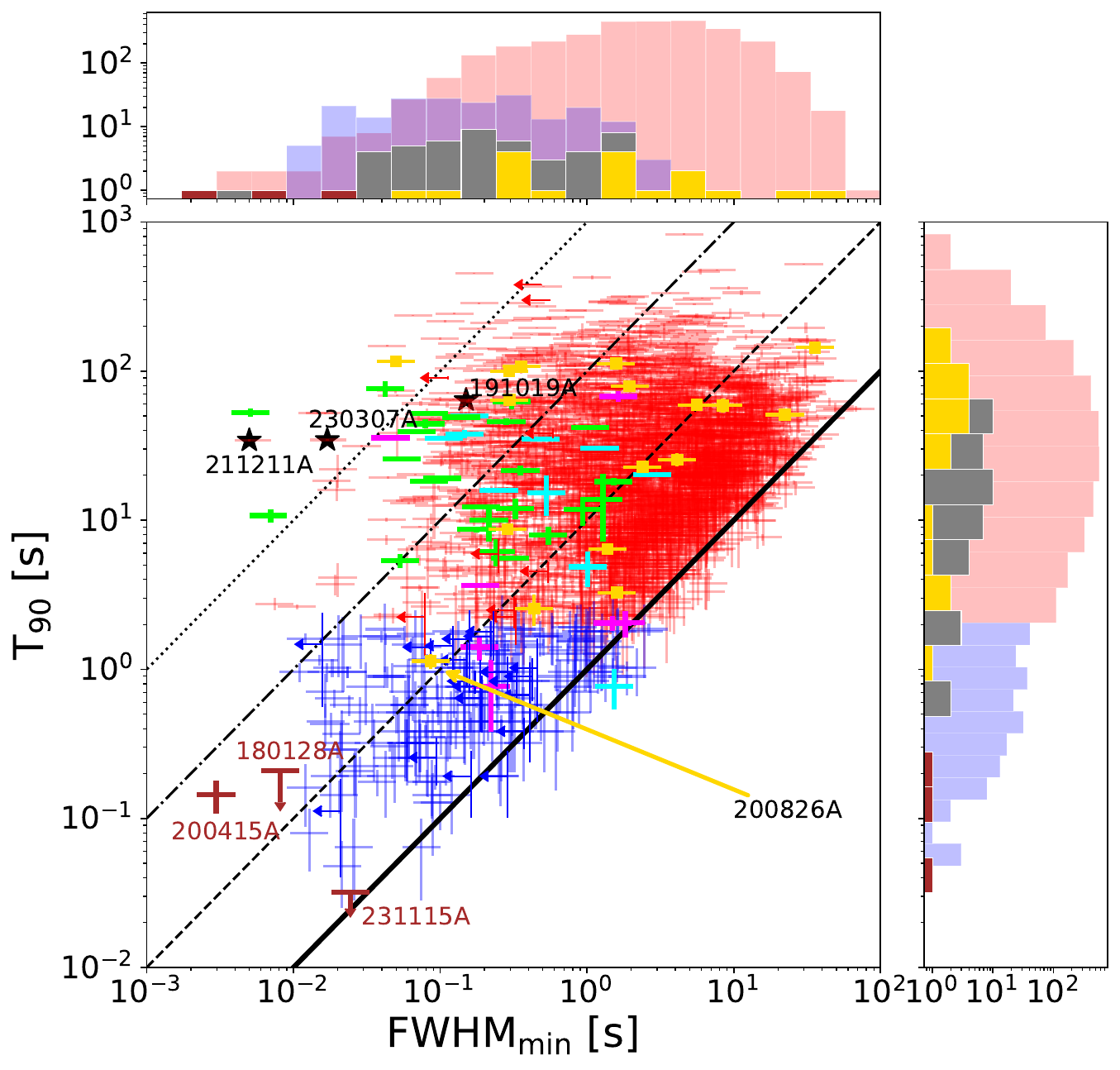}
    \caption{Scatter plot of $\rm{FWHM}_{\rm{min}}$ and $\rm{T}_{\rm{90}}$ for the \textit{Fermi}/GBM sample, along with the corresponding marginal distributions. Blue (red) points represent short (long) GRBs. Gold points represents SN-associated GRBs. Magenta, lime, and cyan points represent SEE-GRBs from \citet{Lien16,Lan20,Kaneko15}, respectively. Three extragalactic MGFs candidates, 180128A, 200415A, and 231115A are shown in brown. SEE-GRBs from the three samples considered are shown altogether in grey in the top and right panel. We also showed with a black star the two peculiar LGRBs 211211A and 230307A associated with a KN event, and 191019A wich can be a short GRB that exploded in a dense environment. We also highlighted the peculiar short collapsar GRB 200826A associated with a SN.} 
    \label{fig:t90_FWHM}
\end{figure*}

Walking in \citetalias{Camisasca23} footsteps, in Figure~\ref{fig:t90_FWHM} we plotted $\rm{FWHM}_{\rm{min}}$ vs. $T_{90}$ for the bursts of our sample. We also computed the median value of $\rm{FWHM}_{\rm{min}}$ for the various GRB classes and applied two-population Kolmogorov-Smirnov (KS) tests to investigate their mutual compatibility. Results are reported in Table~\ref{tab:classif}.

Clearly, LGRBs show greater $\rm{FWHM_{min}}$ values, with a median value of $2.4~\rm{s}$, while it is about $0.2~\rm{s}$ for the SGRBs. GRBs with an ascertained SN association have typical $\rm{FWHM}_{\rm{min}}$ values of the bulk of LGRBs, with a median value of about $1.6$~s, pointing towards a common collapsar origin for the bulk of LGRBs. 
 
Conversely, SEE-GRBs $\rm{FWHM}_{\rm{min}}$ values (about $0.15~\rm{s}$) are closer to those of SGRBs, rather than LGRBs, supporting a common origin. Our results are consistent with those of \citet{Kaneko15} and \citet{Lan20}. Among them, 161129A was also noted by \citet{Guidorzi24b} to have a combination of high variability $V\sim 0.6$, relatively low luminosity ($L_p\sim 2 \times 10^{51}~\rm{erg~s^{-1}})$, and short MVT, potentially characteristic of long-duration merger candidates. The initial spike has a MVT of about 20 ms, although it is slightly below the threshold detection of our technique, having $\rm{SNR_{\rm{4ms}}}=6.6<SNR_{\rm{4ms}}^{thr} = 6.8$ for the 4 ms binned LC.

\begin{figure*}[h]
    \centering
     \begin{minipage}[b]{0.43\textwidth}
        \centering
        \includegraphics[width=\textwidth]{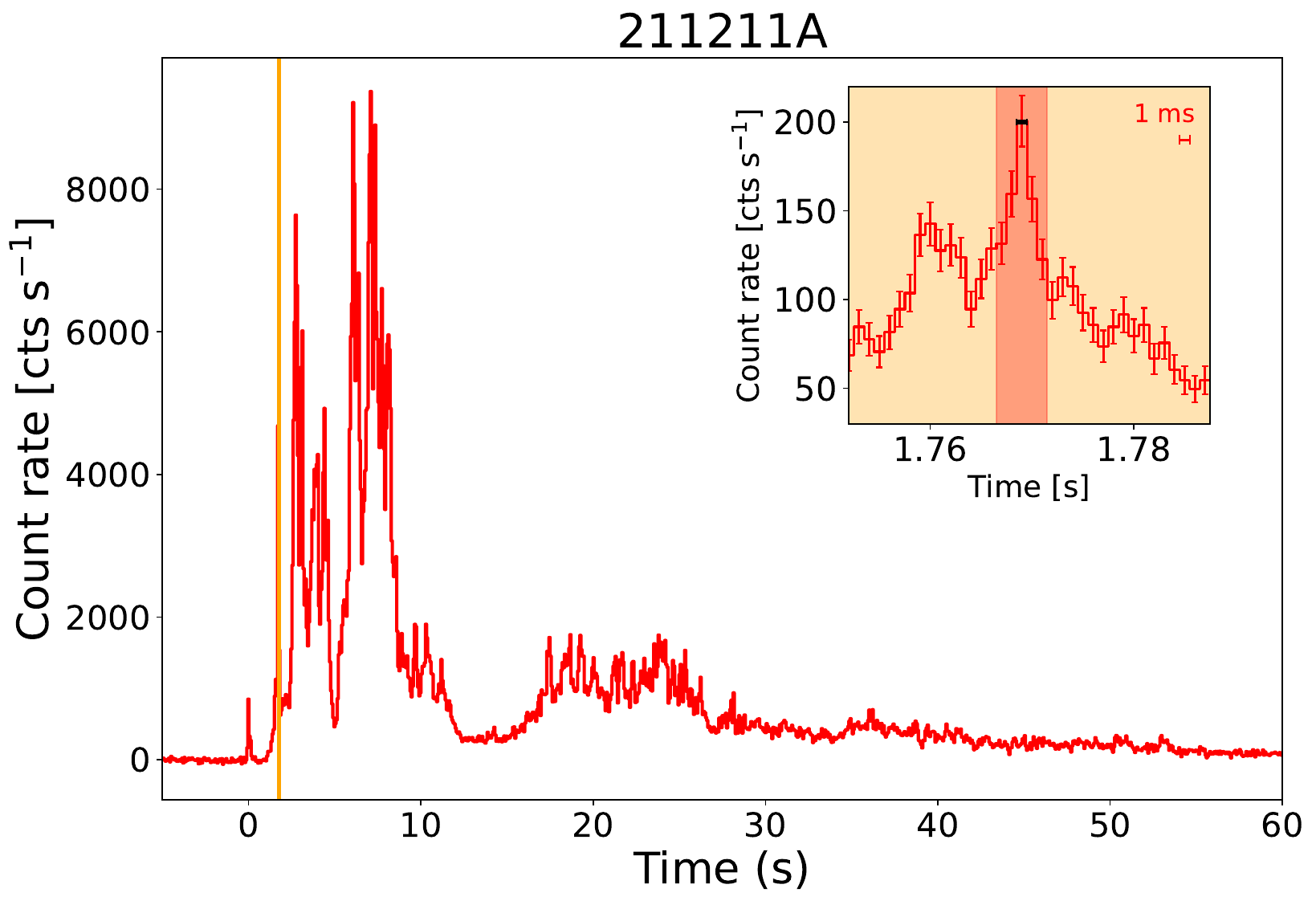}
    \end{minipage}
    \begin{minipage}[b]{0.43\textwidth}
        \centering
        \includegraphics[width=\textwidth]{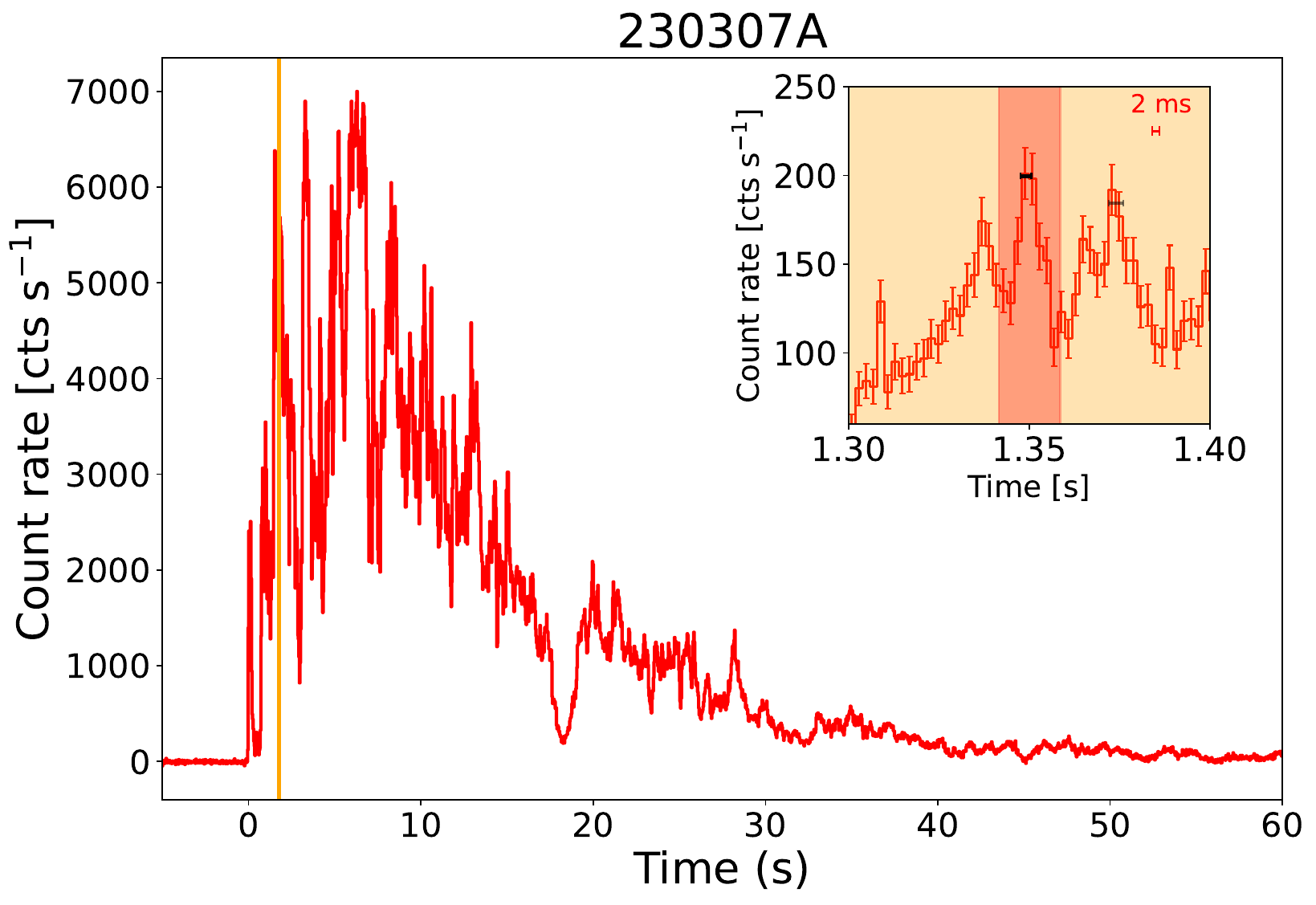}
    \end{minipage}
    \begin{minipage}[b]{0.31\textwidth}
        \centering
        \includegraphics[width=\textwidth]{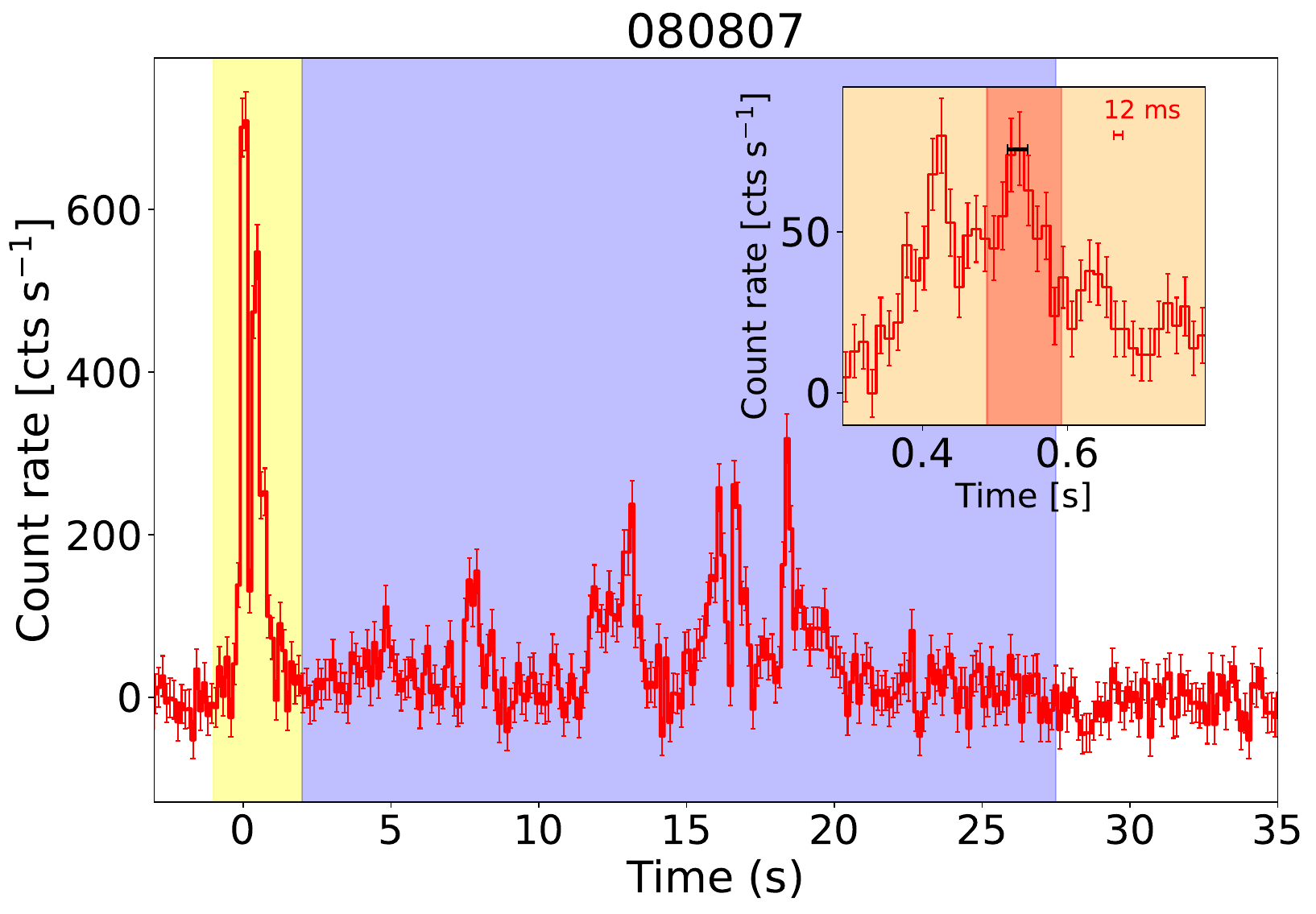}
    \end{minipage}
    \begin{minipage}[b]{0.31\textwidth}
        \centering
        \includegraphics[width=\textwidth]{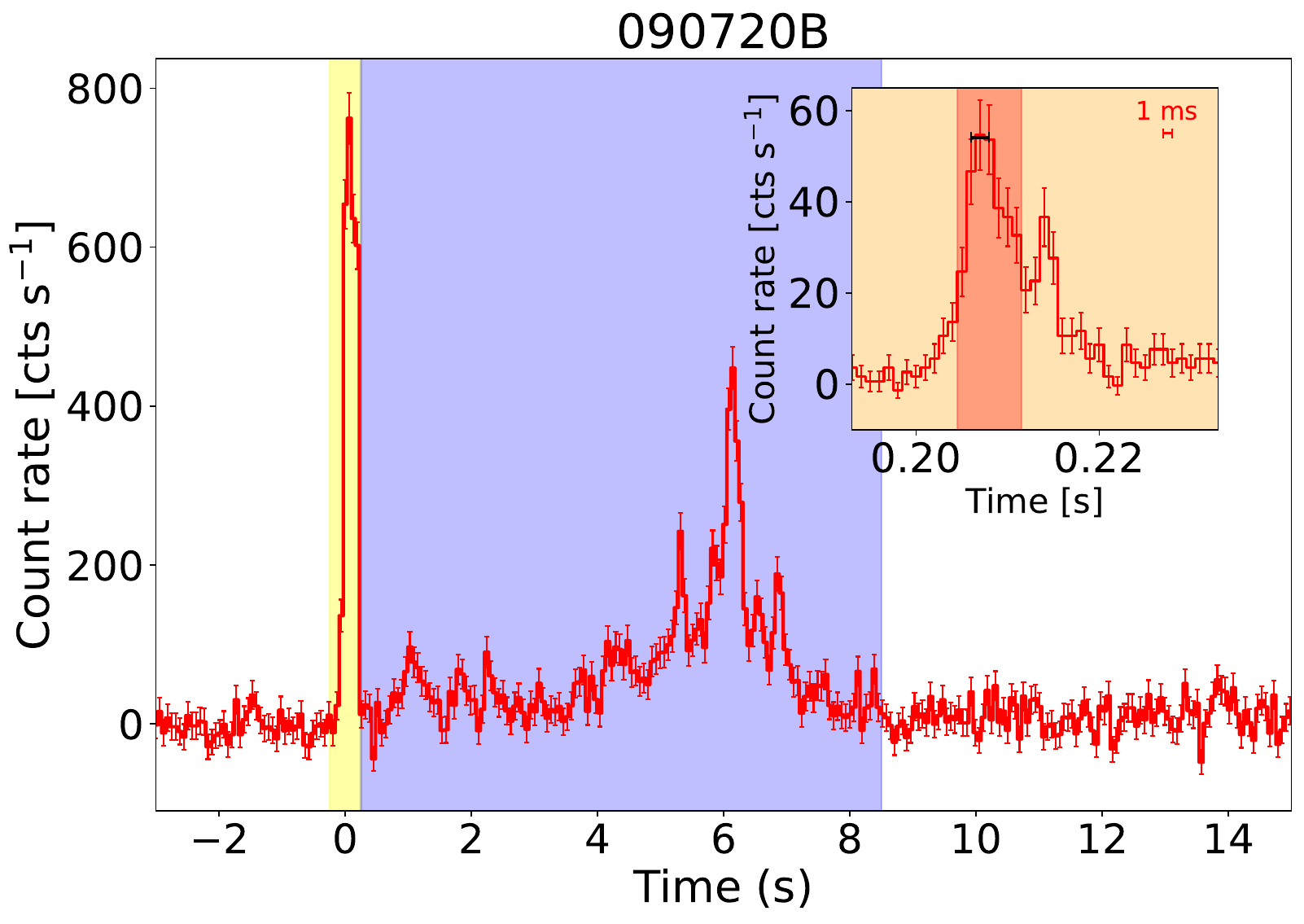}
    \end{minipage}
    \begin{minipage}[b]{0.31\textwidth}
        \centering
        \includegraphics[width=\textwidth]{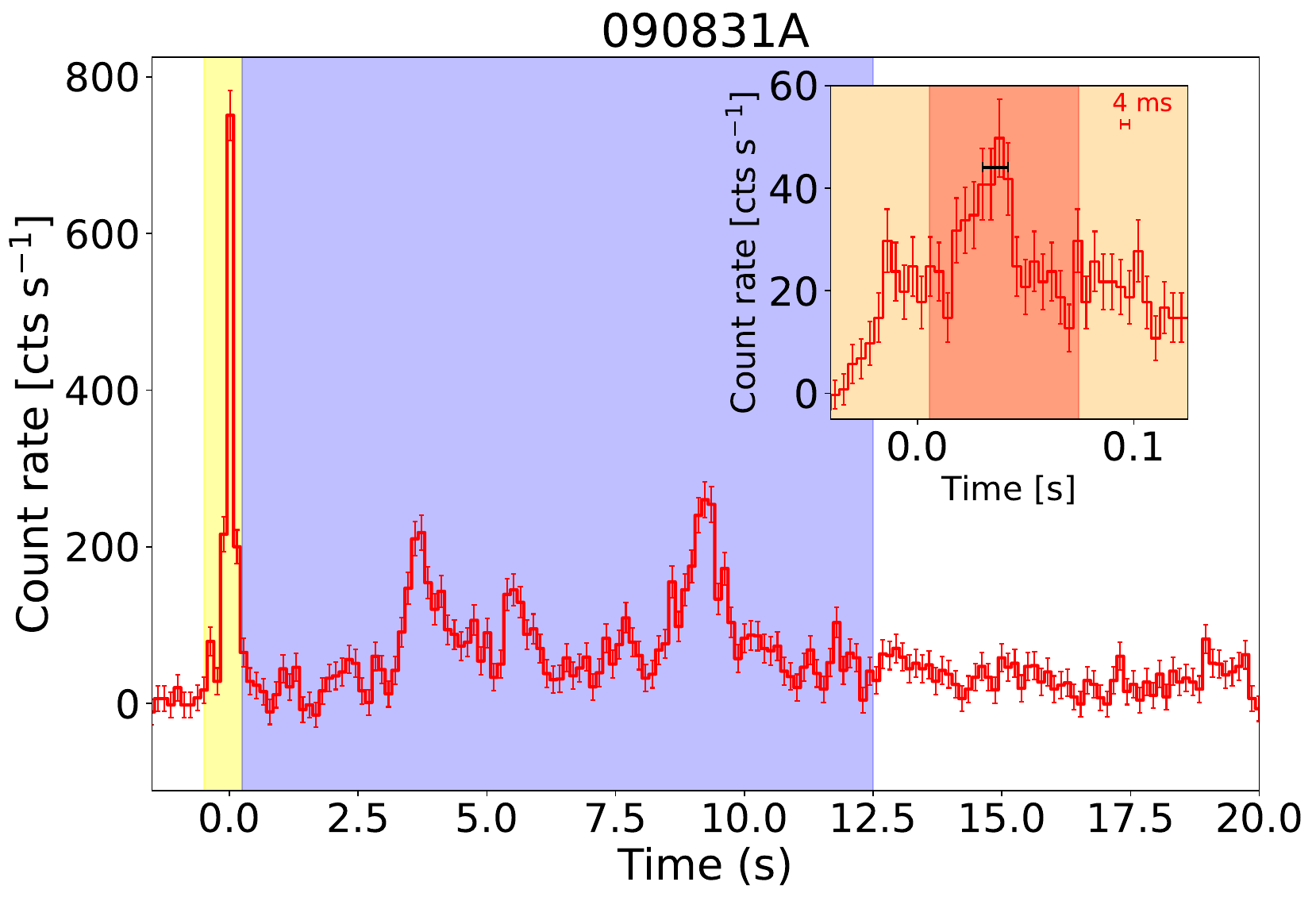}
    \end{minipage}
    \caption{{\it Top panels:} LC of 211211A (left) and of 230307A (right), using the 8-1000 keV range. {\it Bottom panels, left to right:} LC of 080807, 090720B, and 090832A, respectively (same energy range as top panels). The yellow window includes the initial short spike, while the blue one includes the extended emission. The inset of each panel is a close-in on the narrowest pulse. The black point indicates the detection timescale $\Delta t_{\rm det}$ of the narrowest pulse, while the orange region shows the window encompassing $\rm{FWHM_{min}}$.}
    \label{fig:see_grbs}
\end{figure*}

Long-duration merger candidates, like 211211A and 230307A, are definite outliers in the LGRB $\rm{FWHM}_{\rm{min}}$ distribution, with a MVT of $5$ and $17$~ms, respectively. These two cases show that duration alone could be misleading and showcase the potential of MVT to unveil these baffling merger candidates, as already pointed out in \citetalias{Camisasca23} and \citetalias{Veres23}. 
Notably, $2\%$ of LGRBs have $\rm{FWHM_{min}} \leq 0.1~\rm{s}$, indicating that LGRBs could include unidentified merger candidates. A few of these events, which also look like canonical SEE-GRBs, are displayed in Figure ~\ref{fig:see_grbs}. Additionally, we considered 191019A, a long GRB ($T_{\rm{90}}=64~\rm{s}$) at redshift $z=0.248$ and with no associated SN, which might also be a merger candidate \citep{Levan23,Stratta25}. 191019A  was not detected by GBM; however, using {\it Swift}/BAT data \citet{Camisasca23} reported a MVT of $0.196~\rm{s}$. Assuming the scaling of $\rm{FWHM_{min}}$ with photon energy (see \cref{sec:appendix_c}), we estimated that we would have found $0.14-0.15~\rm{s}$ with the GBM, placing 191019A in the outskirts of the LGRB $\rm{FWHM_{min}}$ distribution. 
 
MGFs have a mean $\rm{FWHM}_{\rm{min}}$ of $12~\rm{ms}$, hence exhibiting even shorter values than typical SGRBs. Table~\ref{tab:classif} shows the median $\rm{FWHM}_{\rm{min}}$ value for the different populations as well as the result of the KS tests.

\begin{table}[h!]
\centering
\caption{Median FWHM$_{\rm min}$ values for different GRB groups along with the $p-$values of the two-population KS test between the FWHM$_{\rm min}$ values of each corresponding pair of groups.}
\label{tab:classif}
\begin{tabular}{lccc}
\hline
Sample & $\rm{FWHM_{min}^{\mathrm{(a)}}}$ & LGRBs & SGRBs \\
 & (s)  &  &  \\
 \hline
LGRBs (2994) & 2.4 & - & $10^{-69}$ (\redcross)\\ 
SN GRBs (17) & 1.6& 0.17 (\greencheck) & $7.1~ 10^{-5} $ (\redcross) \\ 
SGRBs (358)& 0.15&$10^{-69}$ (\redcross) & - \\ 
SEE-GRBs$^{\mathrm{(b)}}$  (6) &  0.11& 0.0004 (\redcross) & 0.73 (\greencheck) \\ 
SEE-GRBs$^{\mathrm{(c)}}$  (16) & 0.16 & $7.5~ 10^{-5} $ (\redcross) & 0.75 (\greencheck) \\ 
SEE-GRBs$^{\mathrm{(d)}}$ (22) & 0.2 & $1.3~ 10^{-12} $ (\redcross) & 0.73 (\greencheck) \\ 
MGFs $^{\mathrm{(e)}}$ (3)& 0.008& $2.3~ 10^{-7} $ (\redcross) & 0.003 (\redcross) \\
\hline
\end{tabular}
\begin{list}{}{}
   \item[$^{\mathrm{(a)}}$]{Median value.}
   \item[$^{\mathrm{(b)}}$]{Identified by \citet{Lien16}.}
   \item[$^{\mathrm{(c)}}$]{Identified by \citet{Kaneko15}.}
   \item[$^{\mathrm{(d)}}$]{Identified by from \citet{Lan20}.}
   \item[$^{\mathrm{(e)}}$]{From extragalactic magnetars.}
\end{list}
\end{table}

\begin{figure}
\centering
\includegraphics[scale=0.335]{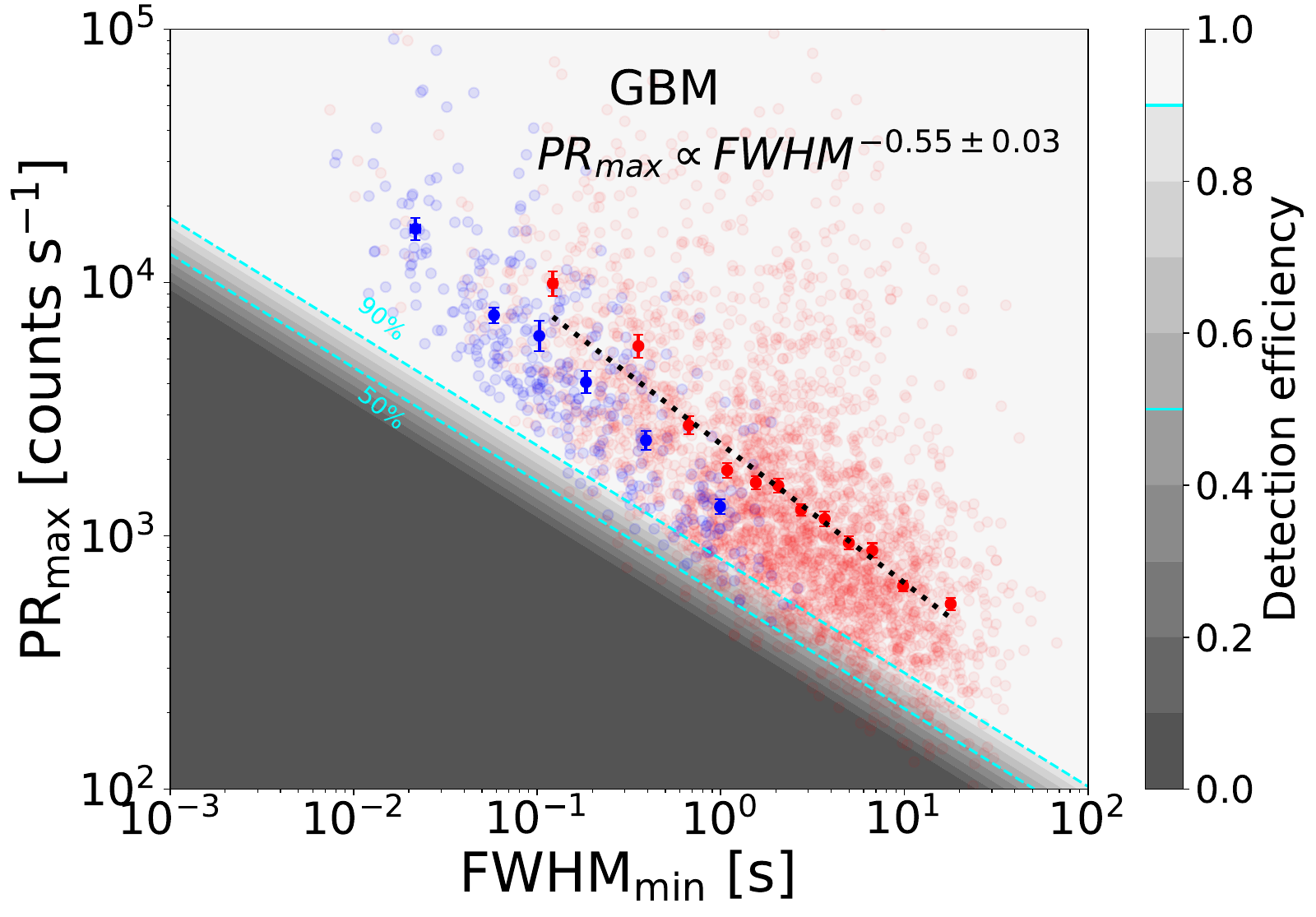}
    \caption{$\rm{PR}_{\rm{max}}$ vs $\rm{FWHM}_{\rm{min}}$ for the GBM sample. Blue dots represent Type-I GRBs (i.e., SGRBs and SEE-GRBs), while red dots represent Type-II GRBs. Lighter dots correspond to individual GRB data, and darker dots indicate the geometric mean of data from GRB groups sorted by increasing $\rm{FWHM_{\rm{min}}}$. Each Type-I group consists of 50 GRBs; each Type-II group consists of 270 GRBs. Dotted lines show the best fit for Type-II GRBs. Shaded areas illustrate ten regions with detection efficiency ranging from 0 to 1. Cyan dashed lines indicate the 50\% and 90\% detection efficiency contours.} 
    \label{fig:FWHM_min_vs_PR_max}
\end{figure}

\subsection{Peak rate versus $\rm{FWHM}_{\rm{min}}$}
\label{sec:FWHM_vs_PRmax}
In line with the procedure of \citetalias{Camisasca23}, we characterised the detection efficiency of {\sc mepsa} applied to GBM data as a function of both $\rm{FWHM}_{\rm{min}}$ and $\rm{PR}_{\rm{max}}$, the latter being the maximum peak rate of any given pulse. To this aim, we generated synthetic pulses assuming the Norris function \citep{Norris96}, and added a constant background with a count rate selected from a sample of real background rates observed with GBM.
For each GRB, the background rate is the sum of the individual rates across all the NaI detectors involved. Poisson noise was finally simulated for the total expected counts per bin. We simulated GRBs with $\rm{PR}_{\rm{max}}$ ranging from $10^2$ to $10^{5}~\rm{cts~s^{-1}}$ and with $\rm{FWHM}_{\rm{min}}$ going from $10^{-2}$ to $10^{2}~\rm{s}$. For each point of this grid, we simulated $100$ pulses and estimated the detection efficiency by counting how many times {\sc mepsa} detected the peak with a $\rm{S/N} > 5$.
 The detection efficiency $\epsilon_{\rm{det}}$ is approximately described by a linear function of the logarithm of both quantities:
 \begin{equation}
 \label{eq:eps_det}
\epsilon_{\rm{det}} = a \log_{\rm{10}}\Bigg( \frac{\rm{FWHM}_{\rm{min}}}{\rm s}  \Bigg)+b\log_{10}\Bigg(\frac{\rm{PR_{\rm{max}}}}{\rm{cts~s^{-1}}}\Bigg)+c.
 \end{equation}
 The optimal coefficients were found to be $a= 1.27$, $b = 2.83$, and $c = -7.33$. Eq.~\eqref{eq:eps_det} is the GBM analogous of eq.~(2) of \citetalias{Camisasca23}:
 \begin{equation}
 \label{eq:eff_det}
     \rm{PR}_{\rm{max}} \geq 877~\rm{cts~s^{-1}}~\biggl(\frac{\rm{FWHM}_{\rm{min}}}{s}\biggl)^{-0.45}10^{0.35(\epsilon-1)}
 \end{equation}
The meaning of Eq.~\eqref{eq:eff_det} is illustrated by the following example: for a pulse with MVT of $10$ ms to be correctly identified with 90\% confidence, its peak rate has to be $\gtrsim 6430$~cts~s$^{-1}$ (a condition that is fulfilled by just 13\% of the bursts in our sample).
Figure~\ref{fig:FWHM_min_vs_PR_max} illustrates
$\epsilon_{\rm{det}}$ in the $\rm{PR_{max}}$--$\rm{FWHM_{min}}$ plane.

\subsection{Peak luminosity versus $\rm{FWHM}_{\rm{min}}$}
\label{sec:lp_mvt}
\begin{figure}
    \centering
\includegraphics[scale=0.25]{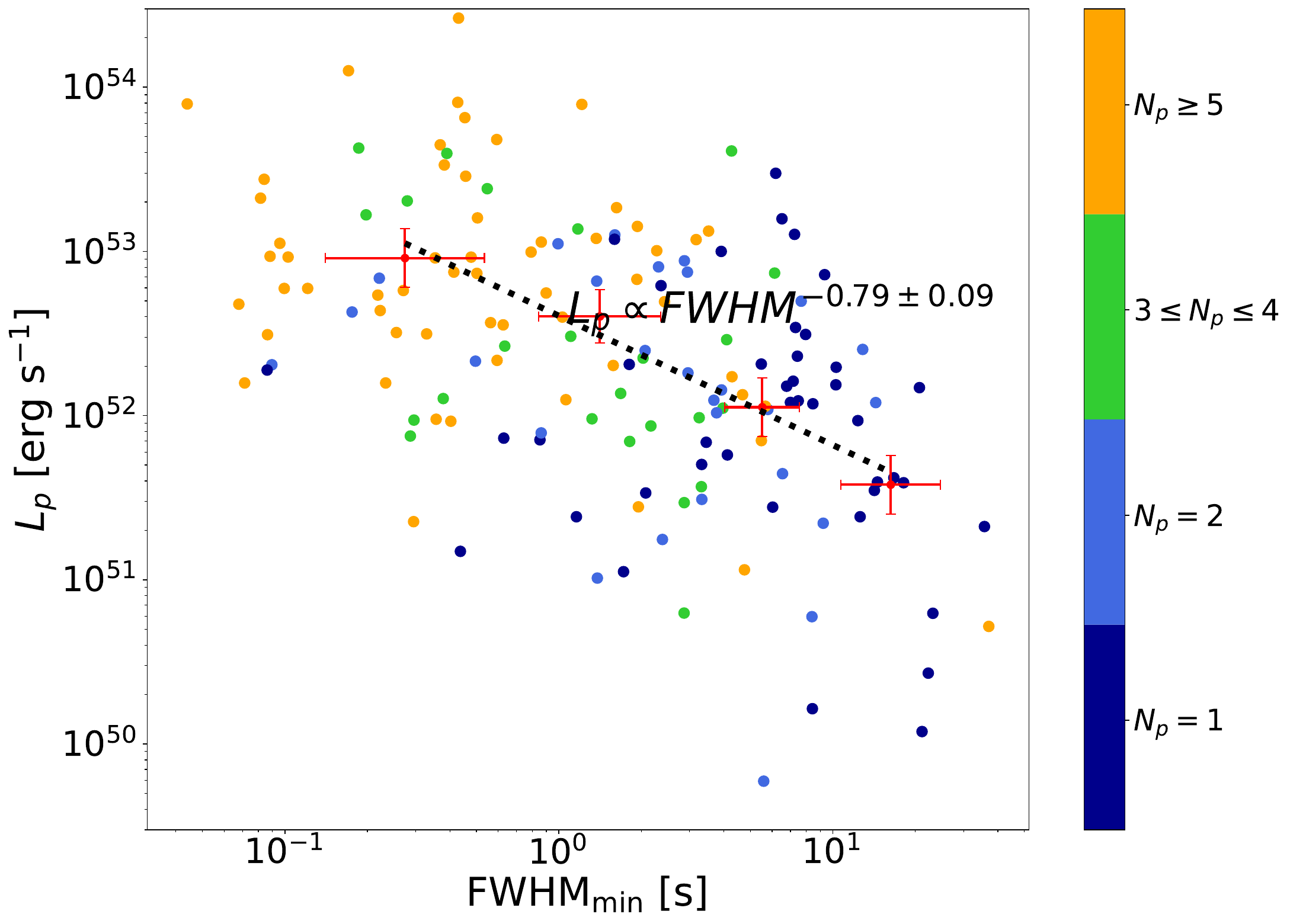}
    \caption{Peak luminosity versus 
    $\rm{FWHM_{min}}$ for collapsar-candidate (or Type-II) GRBs. The red points represent the geometric means of GRB groups sorted by increasing $\rm{FWHM_{min}}$. The dashed line indicates the best fit. GRBs are also categorised by the number of peaks, with the more luminous ones having more peaks.}
    \label{fig:Lpeak_FWHM_Npeaks}
\end{figure}

Isotropic-equivalent peak luminosities $L_{\rm p}$ were computed as in \citet{Maccary24} for 152 collapsar-candidate GRBs with known redshift. 
We studied the $L_{p}-\rm{FWHM}_{\rm{min}}$ correlation, which was observed in other catalogues \citepalias{Camisasca23}. The result is shown in Figure~\ref{fig:Lpeak_FWHM_Npeaks}. 

Selection effects significantly influence the distribution in the $L_p-\rm{FWHM_{min}}$ plane. Specifically, narrower pulses require a higher peak rate to be detected: this selection bias could hide possible weak and short bursts that could contribute to demote the correlation. To account for this bias, we carried out a suite of simulations, following the procedure set up in \citetalias{Camisasca23}. We divided our sample into 9 bins of redshift and simulated points within the $L_p-\rm{FWHM_{min}}$ plane for each bin. For each bin, we randomly generated  as many points as in the corresponding observed sub-sample, where $L_p$ was drawn from the distribution of the observed luminosities in that bin and $\rm{FWHM_{min}}$ was sampled from a probability density function derived from Gaussian kernel density estimation of the Type-II LGRBs with known redshift. Each point was accepted or rejected based on two conditions: (1) a Bernoulli trial with probability $p=\epsilon_{\rm{det}}$ calculated using Eq.~\eqref{eq:eps_det} for that specific point was successful, and (2) the isotropic energy of this synthetic pulse did not exceed the maximum observed energy in that bin, given by $L_p\rm{FWHM_{min}} \leq E_{\rm{iso,max}}^{\rm{(pulse)}}$. 

\begin{figure*}[h!]
\centering
\includegraphics[scale=0.75]{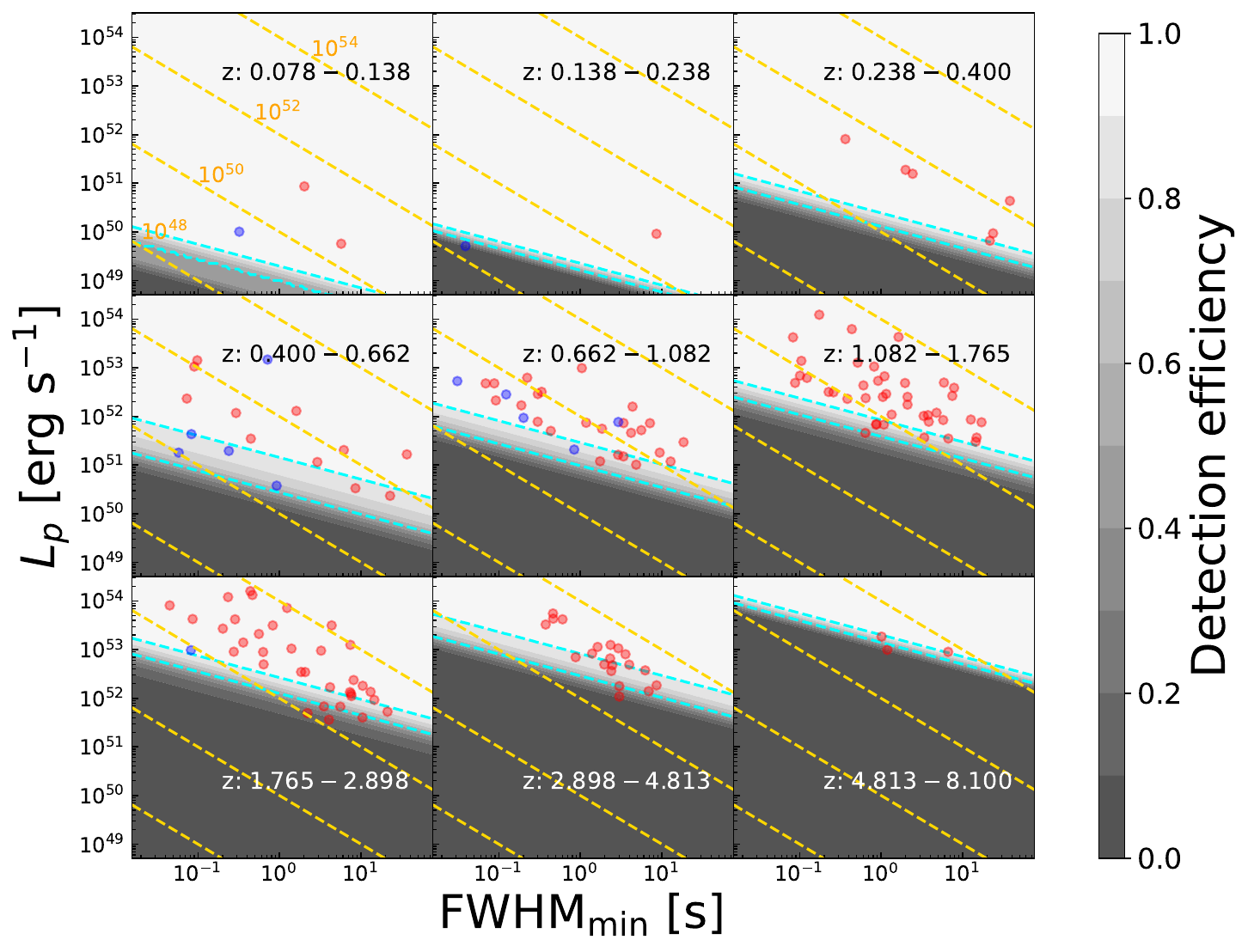}
    \caption{ $L_p$ versus $\rm{FWHM}_{\rm{min}}$ for the \textit{Fermi}/GBM  divided into nine redshift bins with equal logarithmic spacing in luminosity distance. Blue dots represent merger-candidates (or Type-I GRBs), and red dots represent collapsar-candidates (or Type-II GRBs). The dashed cyan lines show 90\% and 50\% detection efficiency (vertical bars). Gold dashed lines indicate regions of constant isotropic-equivalent released energy (in erg) for each peak, roughly calculated as $E_{\rm{iso}} = L_{p} \times \rm{FWHM_{min}}$.}
    \label{fig:Liso_FWHM_simu}
\end{figure*}

We carried out $N=10^{4}$ simulations. This way, we do not assume any correlation between $L_p$ and $\rm{FWHM_{min}}$, whereas the resulting apparent correlation is entirely due to the selection effects (Figure \ref{fig:Liso_FWHM_simu}).

\begin{figure}[h!]
    \centering
\includegraphics[scale=0.24]{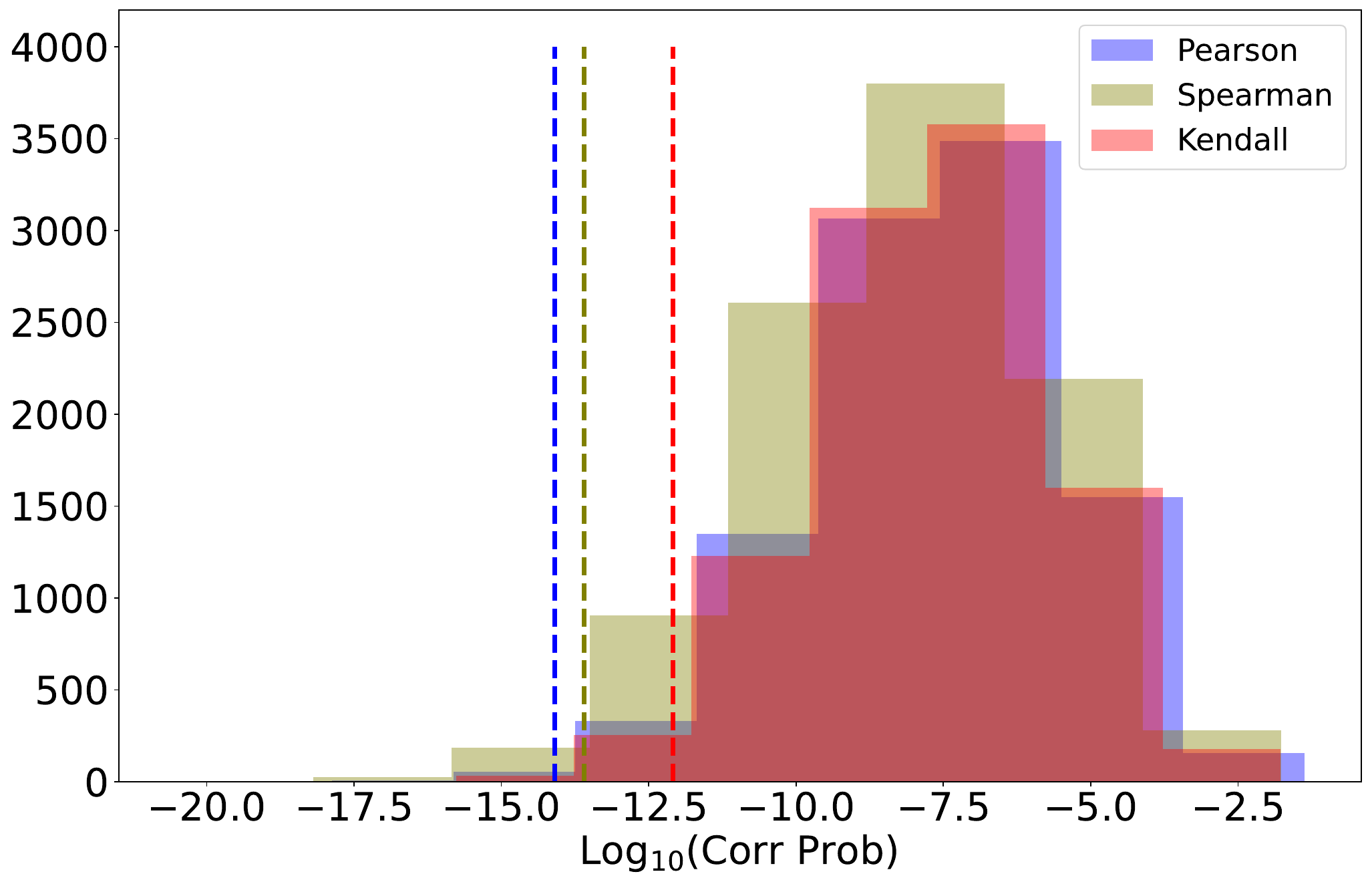}
    \caption{Pearson, Spearman, and Kendall correlation $p$-values (in logarithm units) computed on $N=10^{4}$ simulated samples (blue, olive, and red histograms) compared to the ones computed on the real dataset (blue, olive, and red dashed lines, respectively).}
    \label{fig:correlation_test}
\end{figure}

We then applied Pearson's, Spearman's, and Kendall's correlation tests to the real data, using logarithmic values for the analysis. For Type-II GRBs, we obtained $p$-values of $8\times10^{-15}$, $2.5\times10^{-14}$, and $8\times10^{-13}$, respectively. In contrast, we found no evidence of such correlation for Type-I GRBs, with $p$-values of $0.96$, $0.83$, $0.86$, respectively. We then applied the same correlation test to the $N=10^{4}$ simulated datasets to build the corresponding reference distributions for the $p$-values, that account for the selection effects discussed above. The results are shown on Figure~\ref{fig:correlation_test} and show that the simulated datasets were more correlated than the real ones only in $0.5$\%, $2.1$\%, and $2.3$\% cases, respectively,
which represent the probabilities that the observed correlation could arise purely from selection effects.

\subsection{Number of peaks versus $\rm{FWHM}_{\rm{min}}$}
\label{sec:FWHM_vs_NP}
\begin{figure}[h]
    \centering    \includegraphics[scale=0.26]{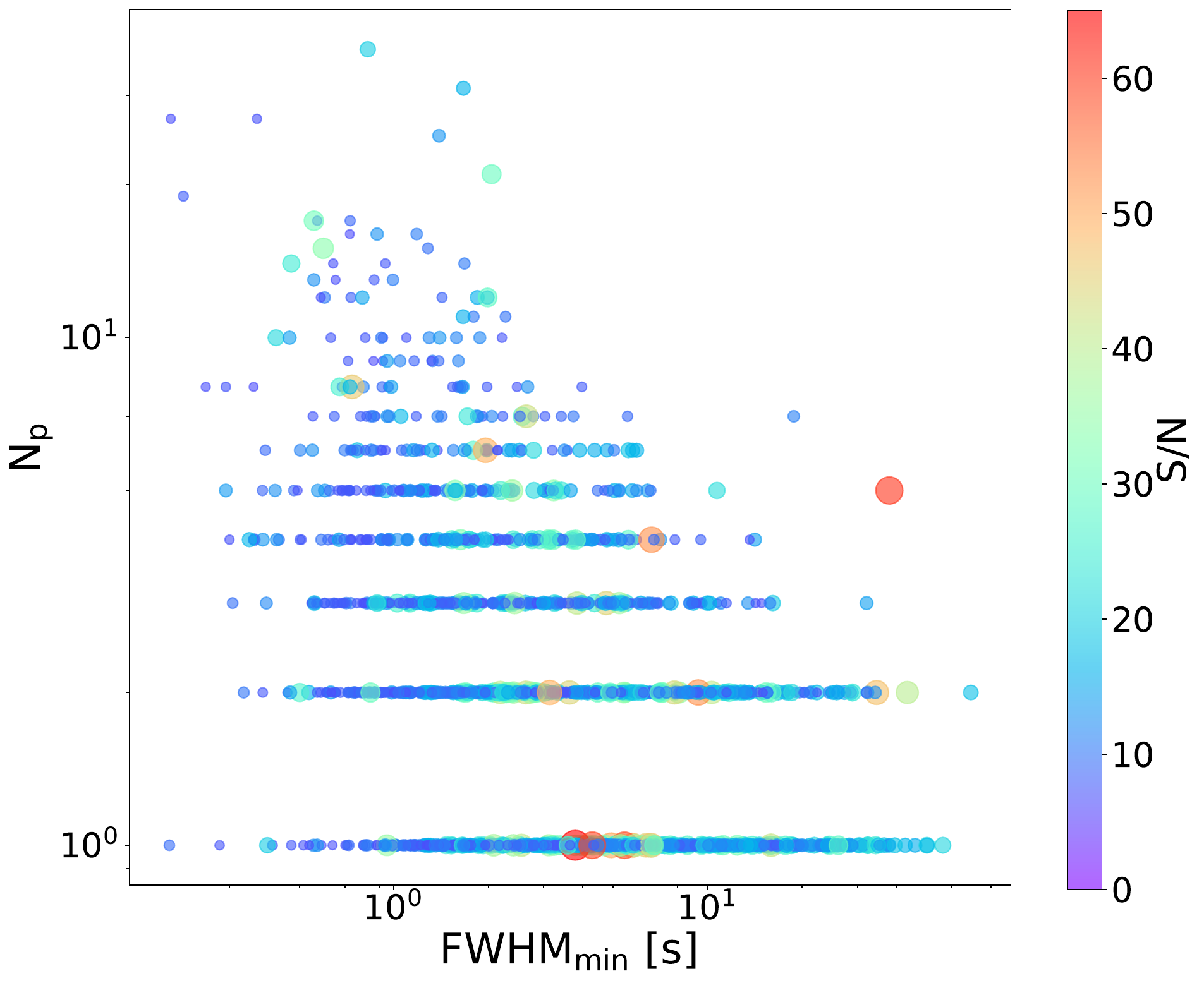}
    \caption{Number of pulses within a GRB as a function of $\rm{FWHM}_{\rm{min}}$, colour coded by S/N. GRBs that are composed of a large number of pulses are more likely to have a shorter $\rm{FWHM}_{\rm{min}}$.}
    \label{fig:npeaks}
\end{figure}

Figure~\ref{fig:npeaks} shows $\rm{FWHM_{min}}$ versus the number of peaks within a GRB. GRBs with numerous peaks (peak-rich GRBs) tend to have shorter MVTs compared with those with fewer peaks (peak-poor GRBs), as was also observed in {\it Swift}/BAT GRBs \citep{Guidorzi16}. The same authors also found that peak-richness correlates with a shallower power density spectrum (PDS), which means that shorter timescales have relatively more temporal power than in peak-poor GRBs \citep{Guidorzi24}.

\subsection{Lorentz factor versus $\rm{FWHM}_{\rm{min}}$}
\label{seq:FWHM_vs_Gamma0}
\begin{figure*}[h]
    \centering
    \begin{minipage}[b]{0.4725\textwidth}
        \centering
        \includegraphics[width=\textwidth]{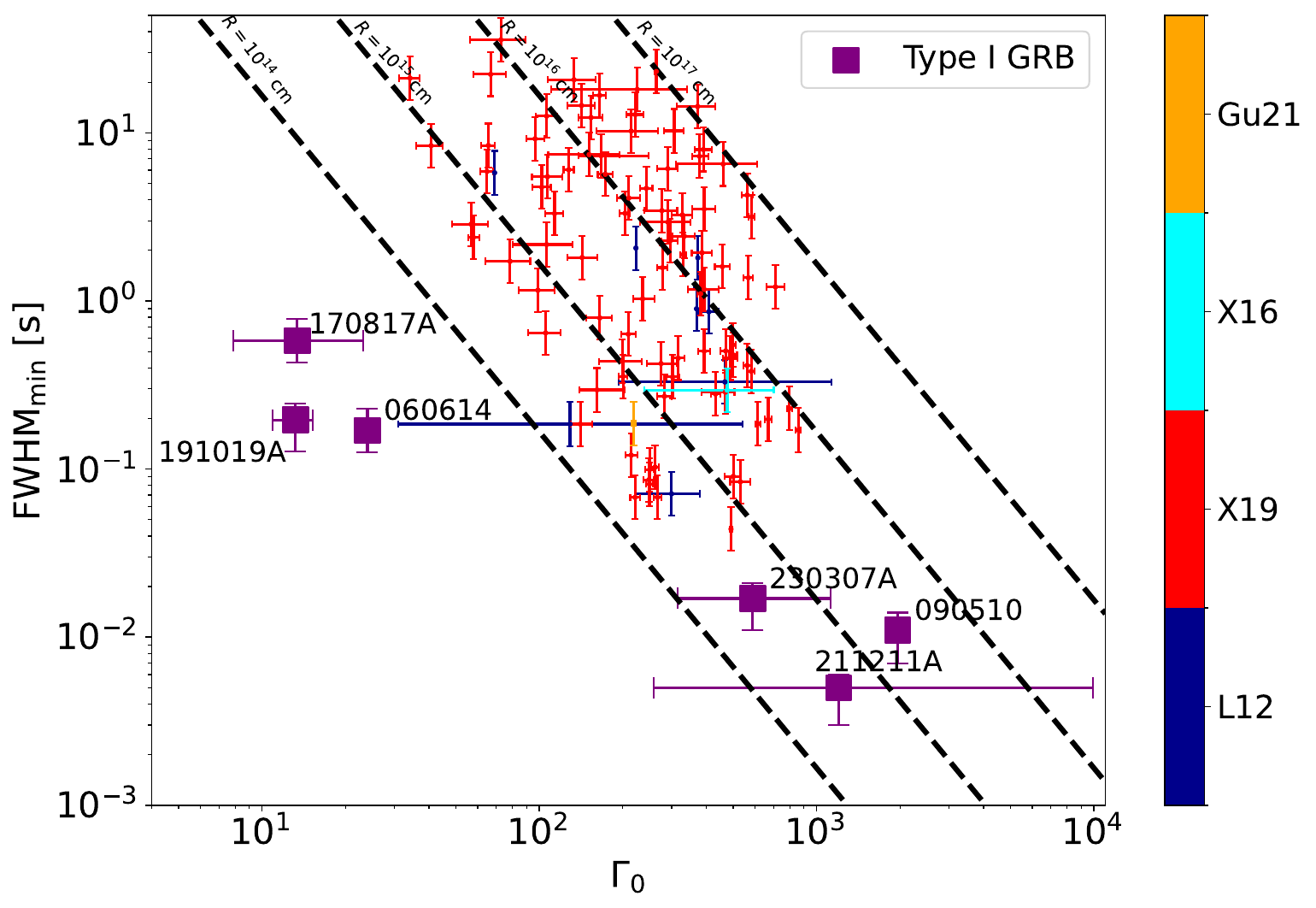}
    \end{minipage}
    \begin{minipage}[b]{0.4\textwidth}
        \centering
        \includegraphics[width=\textwidth]{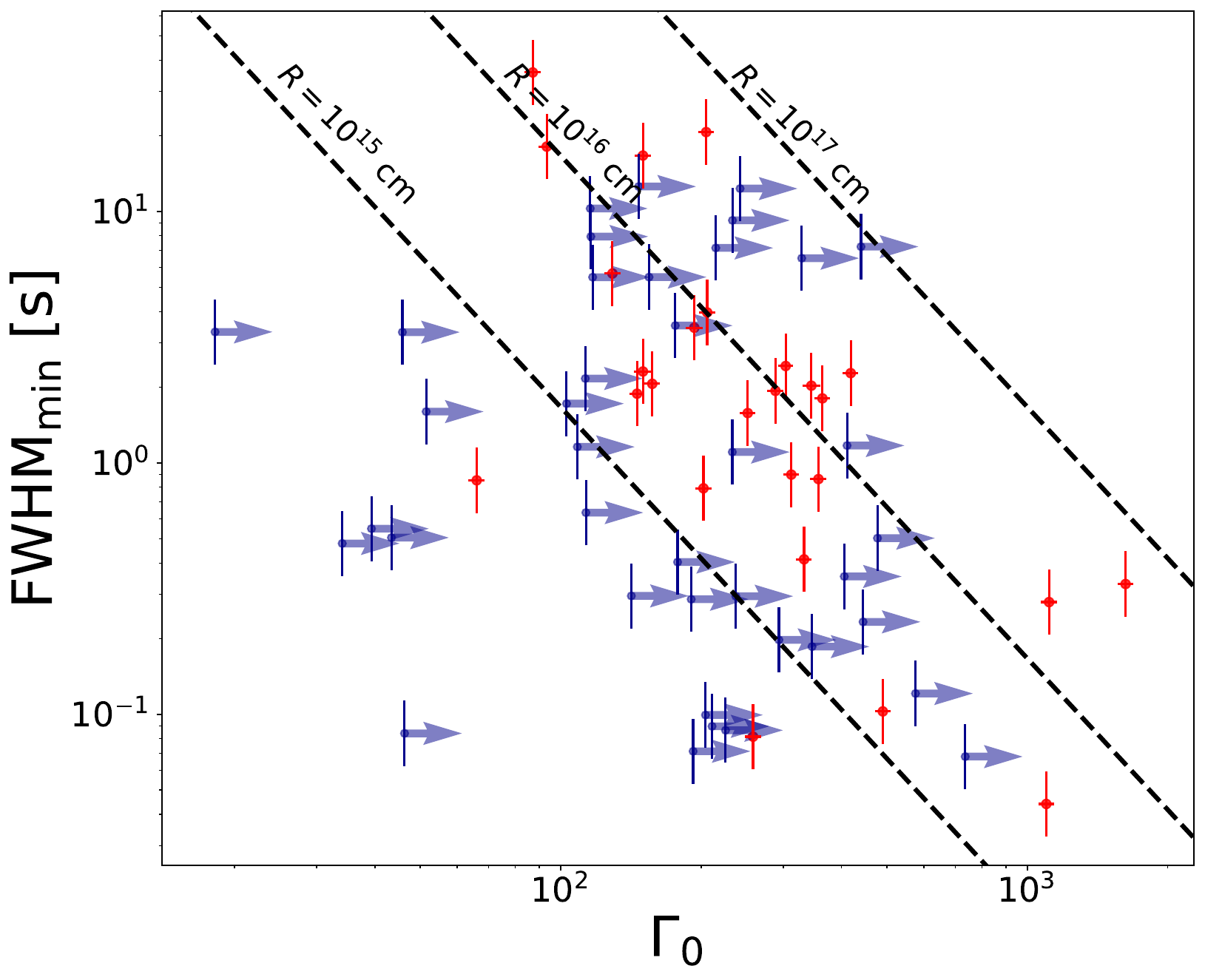}
    \end{minipage}
    \caption{$\rm{FWHM_{min}}$ versus the initial Lorentz factor $\Gamma_{0}$ for the Type-II (red dots) and Type-I GRBs (purple squares) present in our sample}. {\it Left:} the $\Gamma_{0}$ values are taken from different datasets, colour-coded as follows: L12 \citet{Lu12c}, X19 \citet{Xue19b}, X16 \citet{Xin16}, and Gu21 \citet{Gupta21}. The dashed lines represent the typical distance $R=2c\Gamma_{0}^{2}\rm{FWHM_{min}}$ where the dissipation process responsible for the prompt emission could occur. {\it Right:} same as left, but $\Gamma_{0}$ is calculated using the dataset from \citet{Ghirlanda18}. Red points indicate GRBs from their golden and silver samples, while blue points represent lower limits on $\Gamma_{0}$.
    \label{fig:FWHM_min_vs_Gamma_0}
\end{figure*}

We took the Lorentz factors (LFs, hereafter noted as $\Gamma_{0}$) from the same references as in \citetalias{Camisasca23}, that is \citet{Lu12c}, \citet{Xue19b}, and \citet{Xin16}. We found 95 GRBs both detected by \textit{Fermi}/GBM and reported in these studies. For 87 of them, \citet{Xue19b} made use of the $L_{\rm{iso}}-E_{p}-\Gamma_{0}$ correlation to get pseudo values of $\Gamma_{0}$, while for the remaining 9, we used the early afterglow peak to compute $\Gamma_{0}$.
We then considered additional references for individual GRBs: 140102A those $\Gamma_{0}$ was reported by \citet{Gupta21} that modelled the forward and reverse shock, as well as 211211A and 230307A. The LF of 211211A (approximately $\Gamma_0 \sim 1000$) value was obtained either by modelling the forward shock \citep{Mei22}, and by measuring the deceleration peak \citep{Veres23}. According to \citet{Zhong24}, 230307A has $\log(\Gamma_0) \sim 2.77$. Our sample also includes one SGRB,  090510, which is a rare case of SGRB detected by Fermi/LAT. In this case, $\Gamma_0$ was estimated from the peak of the high-energy afterglow \citep{Ghirlanda10b}. The other SGRB in our sample is 170817A, for which a considerably less reliable measure of $\Gamma_0$ was obtained using $E_{\rm p,i}-E_{\rm iso}$ and $\Gamma_0-E_{\rm iso}$ correlations \citep{Zou18}. We also considered the sample of GRBs from \citet{Ghirlanda18} separately. We collected 65 Type-II GRBs in common with our sample, 26 of them being part of their golden or silver sample, and 39 taken as upper limits. We also report the 6 cases of Type-I GRBs for which a measure of $\Gamma_0$ is possible. The results are shown in Figure~\ref{fig:FWHM_min_vs_Gamma_0}. The left panel shows $\rm{FWHM_{min}}$ versus $\Gamma_{0}$, while the right panel displays $\rm{FWHM_{min}}$ versus $\Gamma_{0}$ as measured by \citet{Ghirlanda18}.

\begin{figure}[h]
    \centering
\includegraphics[scale=0.29]{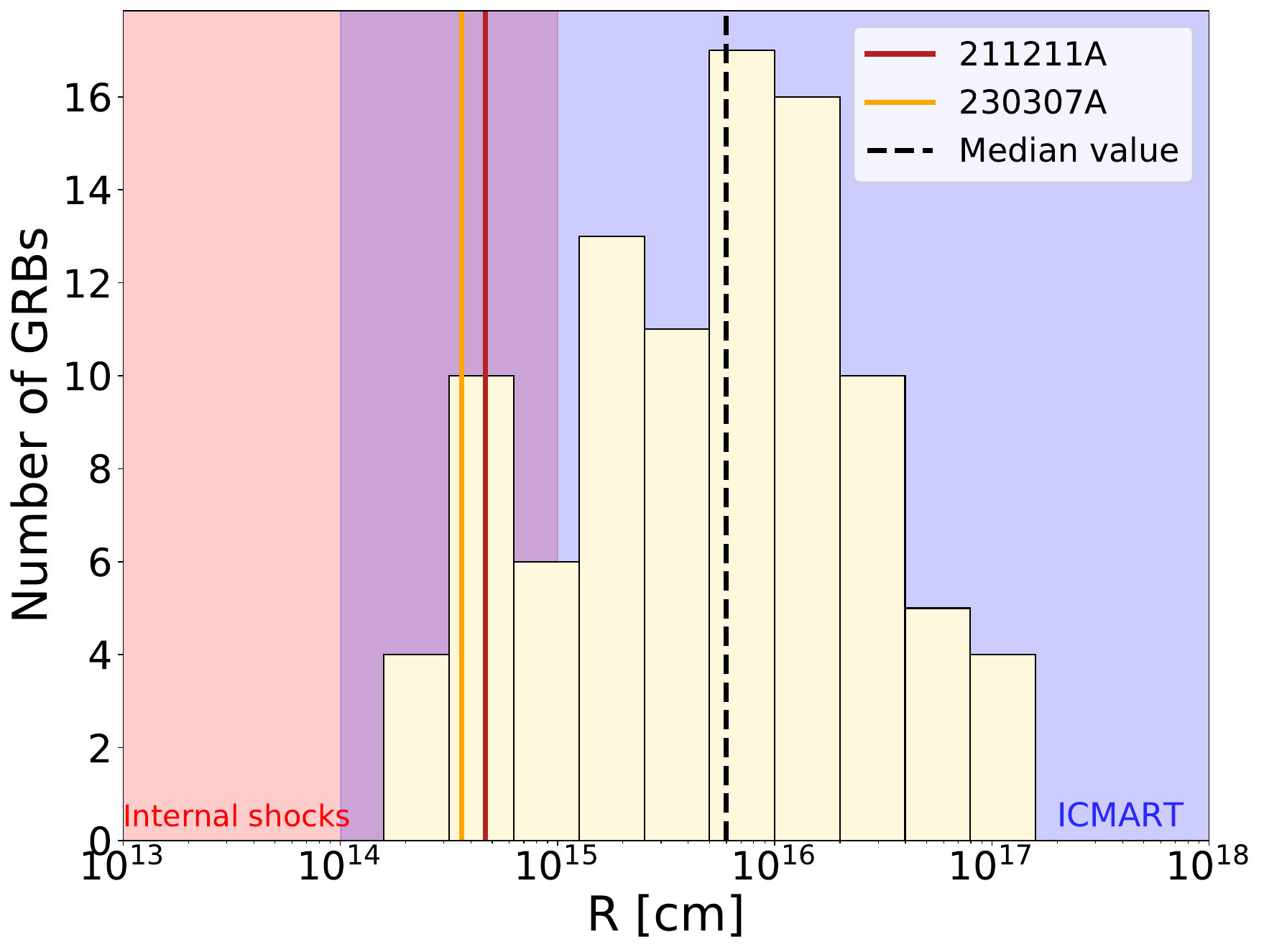}
    \caption{Source emission region radii for the GBM sample. Red and orange vertical lines represents the value of this radius for 211211A and 230307A, while the dashed vertical line represents the median value of the distribution. The red shaded regions indicates the values expected by the IS model ($10^{13}~\rm{cm}\lesssim  R \lesssim 
 10^{15}~\rm{cm}$; \citealt{Rees94,Daigne98}) while the blue one indicate the expectations for the ICMART model ($R\gtrsim 10^{14-15}~\rm{cm}$;  \cite{ICMART}). In the decade $10^{14-15}~{\rm cm}$ (purple), the two regions overlap and the emission radii in this region are marginally compatible with both models.}
    \label{fig:radius}
\end{figure}

We computed the radius $R$ at which the gamma-ray emission is produced using 
\begin{equation}
\label{eq:radius}
    R\ =\ 2\, c\, \Gamma_{0}^{2}\ {\rm FWHM}_{\rm min}\ = 6 \times10^{14}~{\rm cm}~\Bigg [\frac{\Gamma_0}{100}\Bigg]^{2}\Bigg[\frac{{\rm FWHM}_{\rm min}}{1~{\rm s}}\Bigg ].
\end{equation}
Eq.~\ref{eq:radius} derives from the IS model \citep{Rees94,Daigne98}. 
We used ${\rm FWHM_{min}}$ as a proxy of the MVT in the emission radius calculation, rather than ${\rm \Delta t_{min}}$, because the former can be considered a the superposition of the rise and the decay time, while the latter gives only the rise time. This choice is motivated by the fact that the emission radius in the IS framework is linked to the angular spreading timescale, $R/c\Gamma^{2}$, which also governs the decay time of the pulse \citep{Kobayashi02}. Since the decay of GRB pulses is 3-4 times slower than the rise, it is more accurate to consider ${\rm FWHM_{min}}$ than ${\rm \Delta t_{min}}$ when computing the emission radius.
Figure~\ref{fig:radius} shows the $R$ distribution for the Type-II GRBs in our sample. The emission radii $R$ for all our GRBs range from $10^{14}$ to $10^{17}~\rm{cm}$. Notably, $80\%$ of the bursts have $R$ values greater than $10^{15}~\rm{cm}$, with a mean value of $\sim 6 \times 10^{15}~\rm{cm}$. 
As a further test, we computed the deceleration radius $R_{\rm dec}$, and checked that $R<R_{\rm dec}$, using 
\begin{equation}
    R_{\rm dec} = 6.2\times10^{16}\ {\rm cm}\ \ \ E_{{\rm iso},52}^{1/3}\,\Gamma_2^{-2/3}\,n^{-1/3}
\label{eq:r_dec}
\end{equation}
where $E_{\rm{iso},52}=E_{\rm{iso}}/10^{52}$~erg, $\Gamma_2=\Gamma/100$, with $E_{\rm{iso}}$ being the explosion energy ($E_{\rm iso}= E_{\gamma,\rm{iso}}/\eta$, with efficiency $\eta$) and $n$ the medium density.
This was derived from \citet{Sari99,Molinari07} and corresponds to the thin shell case.  We assumed a constant density medium of $n=1~\rm{cm^{-3}}$ and a efficiency of $\eta = 0.2$. We found only 2 cases where $R>R_{\rm dec}$, namely 090423 and 171222A, with $R/R_{\rm dec}$ ratios of 1.1 and 1.3, respectively. Unfortunately, there is no broadband modelling of the afterglow for these GRBs, so we are unable to verify whether our fiducial values are accurate estimates of the explosion energy and medium density in these cases.

\subsection{ Peak energy versus $\rm{FWHM}_{\rm{min}}$}
\label{sec:FWHM_vs_Ep}
\begin{figure}[h]
    \centering
\includegraphics[width=0.47\textwidth]{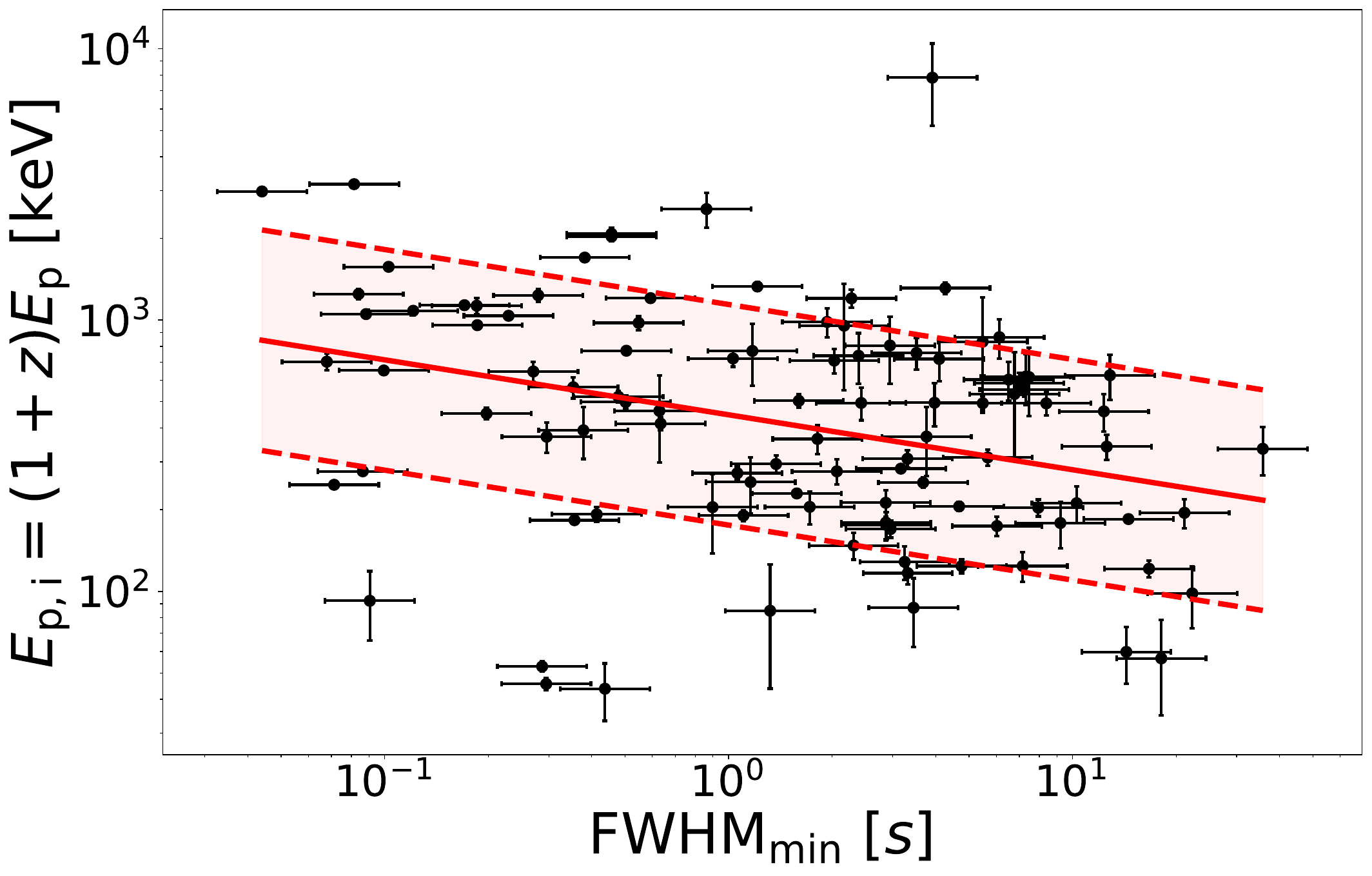}
    \caption{$E_{p,i}$ vs $\rm{FWHM}_{\rm{min}}$ for all Type-II GRBs with known redshift. The solid red line shows the best fit, while the two dashed red lines show the dispersion of the correlation.}
    \label{fig:FWHM_min_vs_Ep}
\end{figure}

We explored the relationship between the MVT and the peak energy $E_p$. We took the $E_p$ information from the GBM catalogue \citep{Goldstein12}. For 1921 bursts, both measures of $E_p$ and FWHM$_{\rm min}$ are available. For a subsample of 107 with measured redshift we could also compute the rest-frame peak energy $E_{p,i} = (1+z) \, E_{p}$. We calculated the Pearson, Spearman, and Kendall correlation coefficients using the logarithmic values, which turned out to be $-0.29$, $-0.30$, and $-0.21$, with associated $p$-values of ($5, 2.4$ and $2) \times 10^{-4}$, respectively). This suggests that the two quantities are somehow correlated, despite the large dispersion, as shown in Figure ~\ref{fig:FWHM_min_vs_Ep}.

We fitted the data with a power-law, $\log{(E_p/{\rm keV})} = m\,\log{({\rm FWHM}_{\rm min}/{\rm s})} + q$, and modelled the dispersion as a further parameter, minimising the D'Agostini likelihood \citep{DAgostini05}. This likelihood is suitable to model correlations affected by a significant scatter, which is treated as a model parameter and is referred to as intrinsic dispersion of the correlation, denoted with $\sigma$.
The resulting parameters are $m=-0.19^{+0.10}_{-0.09}$, $q=2.66\pm 0.07$, and
$\sigma=0.43^{0.06}_{-0.05}$.

\section{Discussion and conclusions}
\label{sec:discussions.}
As previously investigated in \citetalias{Camisasca23}, we confirmed that $\rm{FWHM_{min}}$ is a robust estimate of the MVT, although it carries a slightly different meaning compared to \citetalias{Golkhou15}, which is more tightly connected to the rise time of the pulse. The {\sc mepsa} detection timescale, on the other hand, proved to be also a good indicator of the rise time of the narrowest pulse, providing us a very simple and practical method to compute it. These differences are important to keep in mind, especially when the MVT values are interpreted within a theoretical context.

In the IS model, the hydrodynamic timescale and the angular spreading timescale govern the rise and the decay time of a GRB pulse, respectively (e.g., \citealt{Kobayashi02}). The hydrodynamic timescale, which dictates the rise time, is the shock-crossing time, approximately
$\sim l/c$, where $l$ represents the characteristic irregularity scale or shell width in the outflow. The angular spreading timescale, determining the decay time, results from the time delay or spread due to the angular extent of the emission region. This timescale is roughly
$R/c\Gamma^2$, where $R$ is the emission radius and $\Gamma$ is the Lorentz factor of the emission region. Since most observed pulses exhibit faster rise than they decay, the pulse width is primarily set by the angular spreading time.

We confirmed that millisecond-long pulses are very rare in GRBs. While partly affected by detection thresholds, their scarcity appears to be a genuine feature. Investigating three independent studies, we confirmed that SEE-GRBs have shorter MVTs than LGRBs, compatible with the bulk of SGRBs. Additionally, the three well-known long-duration merger candidates 211211A,  230307A, and 191019A in a lesser extent have very short MVTs (5, 17, and 150 ms) , providing further evidence that short MVTs are characteristic of Type-I GRBs, regardless of the total duration.
Our MVT results align with the conventional interpretation that SEE-GRBs are essentially SGRBs with an additional emission component. The exact physical mechanism behind the extended emission remains unclear. Several models have been proposed, including long-lasting activity from the central engine, such as a magnetar formed during the merger \citep{Metzger08,Jordana22} or energy release from a late fallback accretion disc \citep{Rosswog07,Musolino24}, both of which can continue powering the emission after the main burst.

These results emphasise the importance of multi-wavelength follow-up observations, particularly for LGRBs with low MVTs, as these could reveal other merger events, that might be otherwise misclassified as collapsar candidates.

Differences in MVT in Type-I/Type-II GRBs may indicate distinct progenitors, or disparities in jet propagation.
The irregularity in GRB jets arises from a combination of internal factors, such as variability in the central engine, and external factors like interaction with the surrounding medium and jet instabilities. The conventional central engines are magnetars or black hole accretion disc systems. Even if both types of GRBs are powered by black hole accretion discs, the black holes in Type I events (SGRBs) likely have smaller masses, resulting in shorter dynamical timescales. GRB jets must also penetrate a dense medium surrounding the central engines: a stellar envelope in the case of LGRBs (Type II) or neutron star merger ejecta in the case of SGRBs (Type I). This interaction fosters the growth of hydrodynamic instabilities along the jet boundary (e.g., \citealt{Gottlieb20b}). Type II jets are likely more unstable due to the higher density of the surrounding stellar envelope. Additionally, pulses emitted within the photospheric radius are obscured, adding further complexity.

Extragalactic MGFs have an even shorter MVT than every GRB population, offering us an additional tool to distinguish them from traditional GRB events.

We confirmed, using an independent data set, that in the case of Type-II GRBs, peak luminosity does correlate with $\rm{FWHM_{min}}$, while the same does not hold true for Type-I GRBs: the question as to whether this is a result of a much poorer sample or it is due to the intrinsic absence of correlation will be addressed through future richer data sets.
{We confirmed that GRBs with many pulses (pulse-rich GRBs, as defined in \citealt{Guidorzi24}) tend to have shorter MVTs, supporting the presence of two temporal behaviours: rapid variability atop a slower FRED-like envelope, and purely slow, FRED-like evolution. This distinction may reflect differences in central engine activity, circumburst interactions, or progenitor type.
We computed Lorentz factor and source emission radius $R$.
These $R$ values generally do not align with the IS model, where $R$ typically ranges from $10^{13}$ to $10^{14}~\rm{cm}$.
However, they are consistent with the ICMART model, which predicts gamma-ray emission at larger radii $R>10^{15-16}~\rm{cm}$ \citep{ICMART} through magnetic reconnection cascades.

We investigated the plausible correlation between  MVT and peak energy. Given the established anti-correlation between MVT and peak luminosity (see also \citetalias{Camisasca23}, \citealt{Wu16}), and the known correlation between peak energy and peak luminosity \citep{Yonetoku04,Ghirlanda05}, in principle, we expect an anti-correlation between the MVT and $E_{p,i}$. Moreover, several GRBs with small MVT have also been detected at higher energies by \textit{Fermi}/LAT such as 080916C ($0.3~\rm{s}$, \citealt{Tajima08}), 090510 ($0.011~\rm{s}$, \citealt{Ohno09}), 090720B ($0.014~\rm{s}$, \citealt{Rubtsov12}), 210410A ($0.07~\rm{s}$, \citealt{Arimoto21}). 
 Although we do find a correlation, it is very dispersed. Smaller MVTs may imply shorter angular spreading times and smaller emission radii, resulting in higher shock energy density in the emission region. In the standard synchrotron shock model, a constant fraction of the shock energy is transferred to magnetic fields, with radiation from smaller radii generally expected to be harder. However, since both the shock energy generated through internal dissipation and the blue-shift of emission frequencies depend on the Lorentz factor, velocity irregularities in the outflow—an essential assumption of the IS model—can introduce significant dispersion in this relationship.


 
\section{Data availability}
\label{sec:data_av}
Table~\ref{tab:dataset} is only available in electronic form at the CDS via anonymous ftp to cdsarc.u-strasbg.fr (130.79.128.5) or via \href{http://cdsarc.u-strasbg.fr/viz-bin/qcat?J/A+A/}{http://cdsweb.u-strasbg.fr/cgi-bin/qcat?J/A+A/}.

\begin{acknowledgements}
We are grateful to the anonymous reviewer for their precious report which helped us to cross-check our results and to overall improve the quality of this work.

R.M. and M.M. acknowledge the University of Ferrara for the financial support of their PhD scholarships. L.F. acknowledges support from the AHEAD-2020 Project grant agreement 871158 of the European Union’s Horizon 2020 Program.
 A.T. acknowledges financial support from ASI-INAF Accordo Attuativo HERMES Pathfinder operazioni n. 2022-25-HH.0 and the basic funding program of the Ioffe Institute  FFUG-2024-0002. L.A. acknowledges support from INAF Mini-grant programme 2022. A.E.C received support from the European Research Council (ERC) via the ERC Synergy Grant ECOGAL (grant 855130). Views and opinions expressed by ERC-funded scientists are however those of the author(s) only and do not necessarily reflect those of the European Union or the European Research Council. Neither the European Union nor the granting authority can be held responsible for them. M. B. acknowledges the Department of Physics and Earth Science of the University of Ferrara for the financial support through the FIRD 2024 grant.

\end{acknowledgements}

\bibliographystyle{aa}
\bibliography{alles_grbs}

\FloatBarrier  

\begin{appendix}
\section{$\rm{FWHM_{min}}$ compared to a direct fit of the narrowest pulse}
\label{sec:appendix}

We compared $\rm{FWHM}_{\rm min}$ measurements obtained using {\sc mepsa} calibration and the procedure described in \citetalias{Camisasca23} with the results derived from fitting the LC with FRED shaped pulses (denoted as FWHM$_{\rm fit}$; see \citet{Maccary24} for a detailed description of the technique). To do so, we analysed a sub-sample of GRBs with either one or a few peaks, for which a direct and accurate modelling of the pulses' shapes and FWHMs was feasible. We initially selected 639 single-peaked GRBs with S/N $>10$. We excluded the GRBs that displayed a more complex temporal structure than a single well-shaped pulse, ending up with 544 GRBs. Their pulses were then fitted with a FRED template and discarded the cases, whose best-fit parameters were too close to the boundaries (chosen to avoid unrealistic parameter values), or with relative errors on the rise time greater than 50\%, reducing the sample to 410 GRBs.  We used \textit{Swift}/BAT data as well, taking a sub-sample of GRBs with less than $8$ peaks in their LC. After intersecting these data with the \citetalias{Golkhou15} and \citetalias{Veres23} results, we retained 244 GRBs in the GBM sample and 28 in the BAT sample for a comparative analysis. In~\cref{fig:FWHM_fit_vs_FWHM_min} we illustrated the comparison between $\rm{FWHM}_{\rm min}$ and FWHM$_{\rm fit}$, showing how closely the two methods agree across different GRBs. As we can see, for most peaks, $\frac{\rm FWHM_{\rm fit}}{2}\leq \rm FWHM_{\rm min} \leq 2 \cdot \rm FWHM_{\rm fit}$; more precisely, 90\% of events are in the range $0.66 \cdot \rm FWHM_{\rm fit}\leq \rm FWHM_{\rm min} \leq 1.92 \cdot \rm FWHM_{\rm fit}$. 

We furthermore performed a linear fit of the form $y = mx + q$, applied to the logarithmic values, modelling the intrinsic dispersion $\sigma_{\rm int}$ as a further parameter,  adopting the D’Agostini likelihood \citep{DAgostini05}. Optimising the parameters using MCMC, we found $m=0.990^{+0.011}_{-0.004}$, $q=0.051^{+0.016}_{-0.017}$, and $\sigma_{\rm int}=0.071^{+0.023}_{-0.028}$. The uncertainty on  ${\rm log ~FWHM_{\rm min}}$, previously estimated as $\sigma_{\rm min} = 0.13$ (i.e a $35\%$ relative error on ${\rm FWHM_{\min}}$), leads to a total uncertainty, accounting for the intrinsic dispersion $\sigma_{\rm int}$, of $\sigma_{\rm tot} = \sqrt{\sigma_{\min}^{2} +\sigma_{\rm int}^{2}} \simeq 0.07$ ($41\%)$. This implies that the relative error made when using ${\rm FWHM_{min}}$ instead of ${\rm FWHM_{fit}}$ is approximately $41\%$, as opposed to the $35\%$ estimated on synthetic LCs by \citet{Camisasca23}.
This comparison ensures the robustness of our MVT measurements by validating them against an established LC fitting method.
\begin{figure}[h!]
    \centering
    \includegraphics[width=\linewidth]{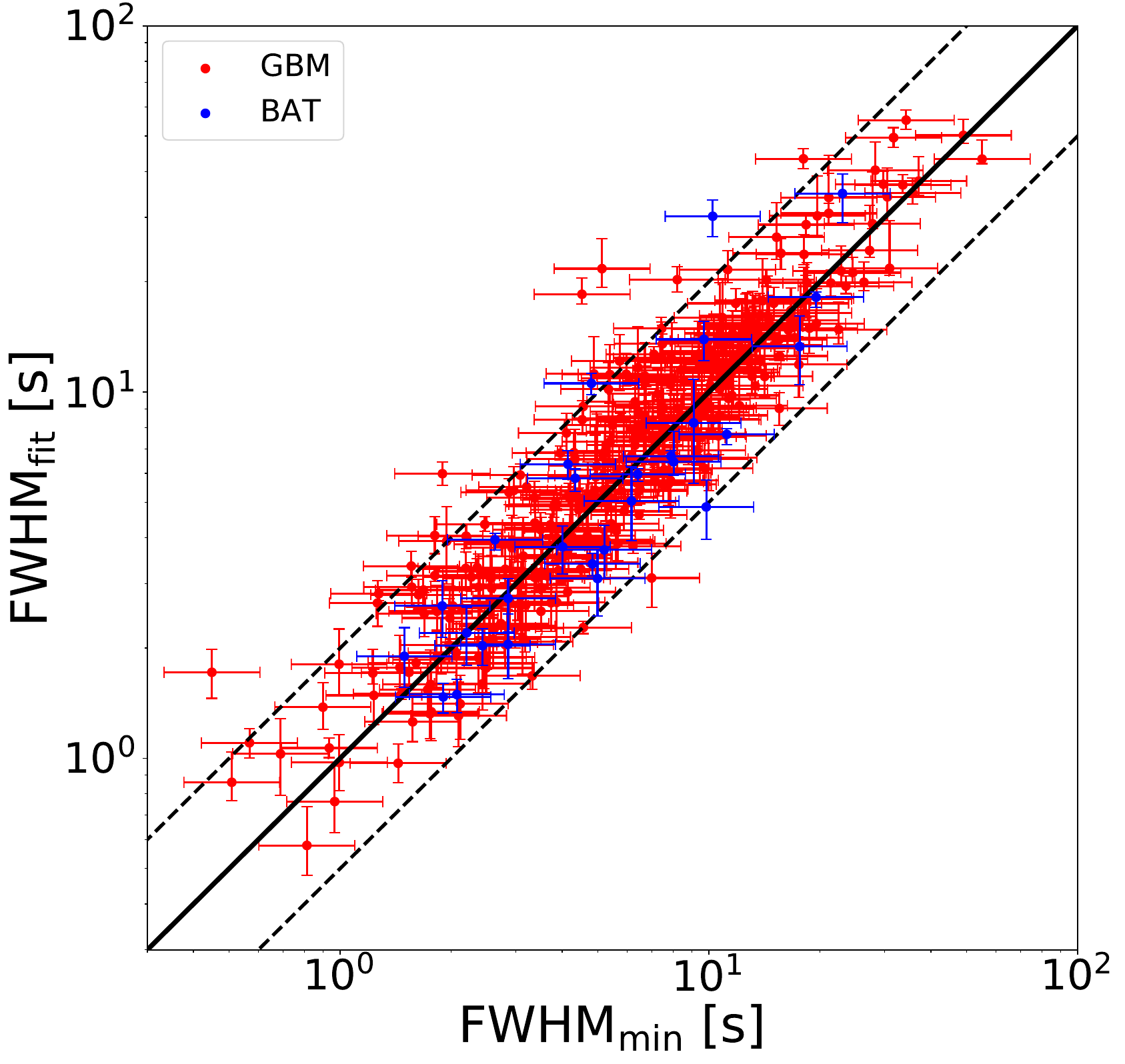}
    \caption{This plot represents $\rm{FWHM_{min}}$, computed either by following the method described in \citetalias{Camisasca23}, or by directly fitting the narrowest pulse by a Norris function (called here $\rm{FWHM_{fit}}$). Red (blue) points were obtained using GBM (resp. BAT) data. The black line indicates the equality line while the dashed lines show a factor 2 of discrepancy, illustrating that most measurements fall within this range.}
    \label{fig:FWHM_fit_vs_FWHM_min}
\end{figure}

\FloatBarrier
\section{Comparison between BAT and GBM}
\label{sec:appendix_B}
We compared the results obtained by \citetalias{Camisasca23} with BAT data with those obtained in this work with GBM data. The $\rm{FWHM_{min}}$ obtained with the GBM is in mean twice as small as those obtained with the BAT. This was expected due to the dependance of the MVT on the energy band.
We also carried this analysis restricting the GBM energy range to the \textit{Swift}/BAT one (15-150 keV). The results show a dispersion around the equality line but no general trend, indicating that our results are consistent with the ones of \citetalias{Camisasca23}. The results of these two analyses are shown in Figure~\ref{fig:swift_fermi}.

\begin{figure*}[h]
    \centering
    \begin{minipage}[b]{0.437\textwidth}
        \centering
        \includegraphics[width=\textwidth]{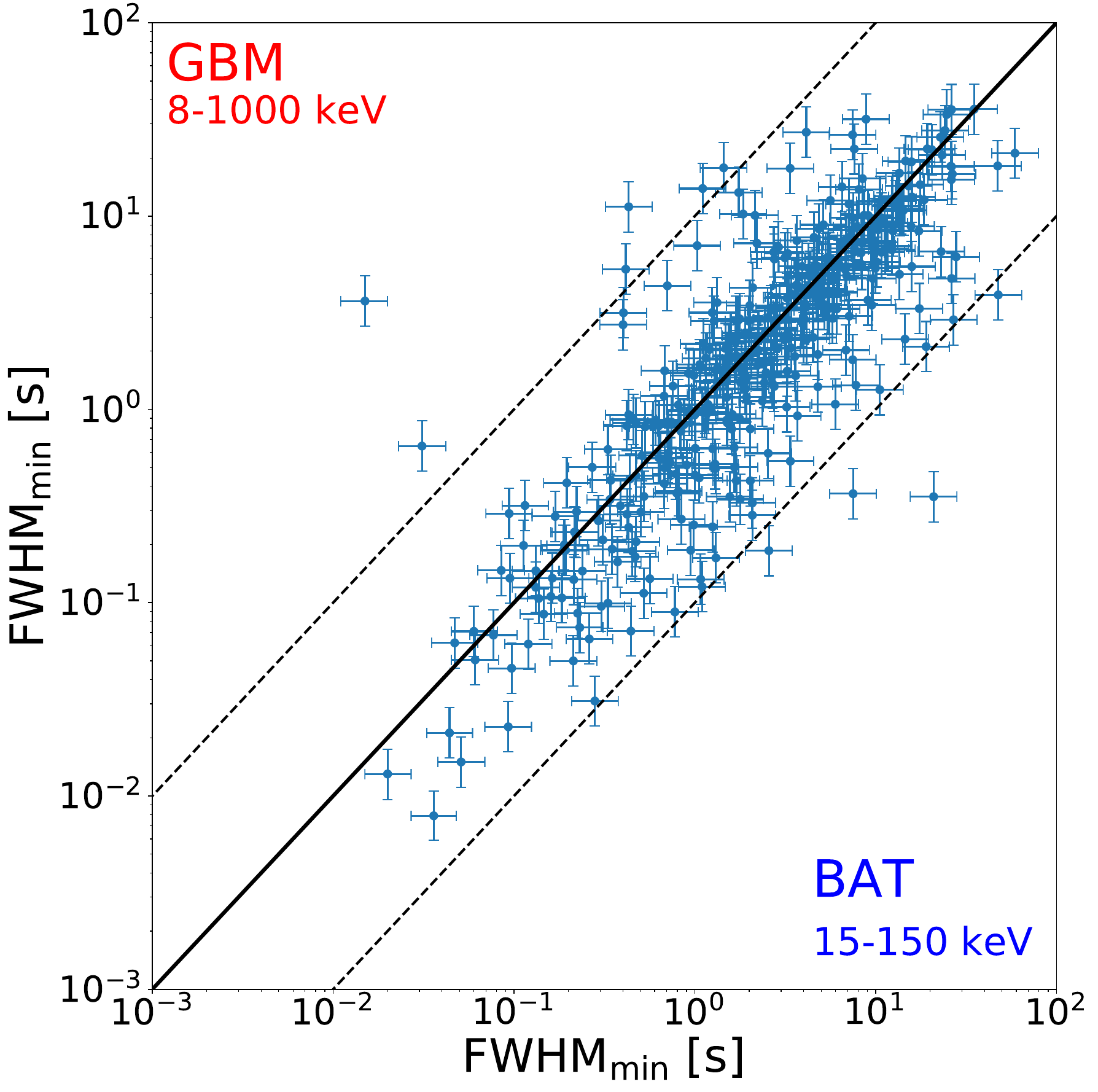}
    \end{minipage}
    \begin{minipage}[b]{0.437\textwidth}
        \centering
        \includegraphics[width=\textwidth]{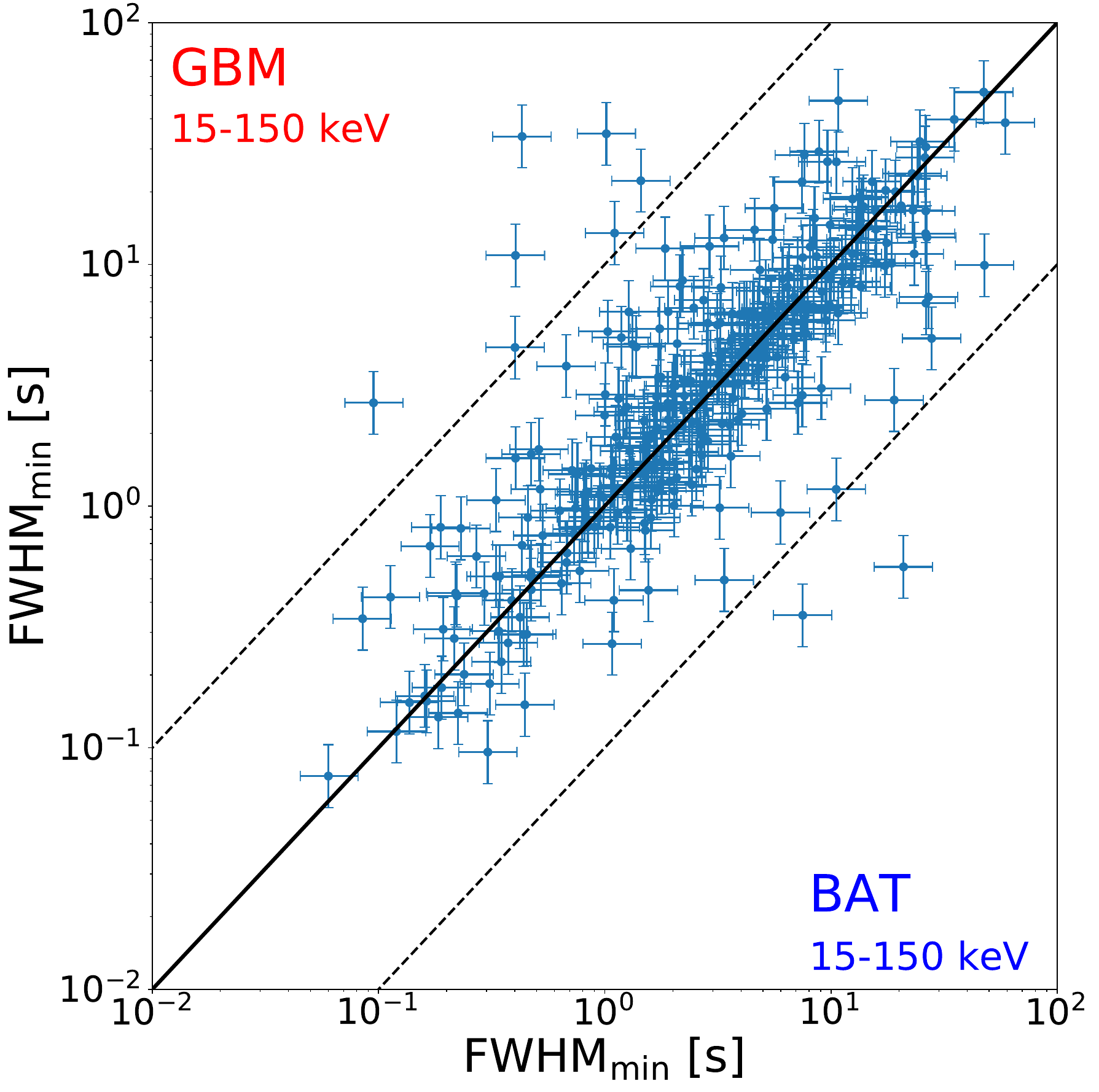}
    \end{minipage}
    \caption{\textbf{Left:}  {$\rm{FWHM}_{\rm min}$ computed by \citetalias{Camisasca23} with BAT data on 15-150 keV against the $\rm{FWHM}_{\rm min}$ of the same bursts but computed in this work on GBM data on the 8-1000 keV range. \textbf{Right}: Same, but the GBM energy range is restricted to 15-150 keV to be the same as BAT.}   }
    \label{fig:swift_fermi}
\end{figure*}

\FloatBarrier

\section{$\rm{FWHM}_{\rm{min}}$ compared with Bayesian blocks}
\label{sec:appendix_bb}
In this section, we compared our results with those obtained by segmenting the LC using Bayesian blocks\footnote{We used the function \textit{bayesian\_blocks} from the \textit{astropy.stats} python library.} (BBs, \citealt{Scargle13}). We applied BBs, using a false alarm threshold of $p_0 = 10^{-3}$ to a sample of 96 GRBs, chosen with $\rm{FWHM}_{\rm{min}} < 50$ ms. This choice was made to obtain enough GRBs to make a sound statistical analysis and to include bright GRBs with evident sub-second structures with exquisite S/N. This sample includes, for instance, GRBs as 211211A, 230307A, 190114C, and others known for their rapid temporal variability and brightness, making them ideal test cases. We computed the MVT using BBs, defining it as the shortest block in the segmentation, ${\rm \Delta T_{BB}}$. 
Figure~\ref{fig:mvt_vs_BBs} compares these values with those from the {\sc mepsa}-based approach. The points scatter around the equality line without a clear systematic bias in either direction. The distribution of the ratio $\Delta T_{BB}/\rm{FWHM}_{\rm{min}} $ is shown in Fig.~\ref{fig:mvts_vs_bbs_ratio} with a median of about $1.03$—meaning that, on average, BBs yield MVT values $\sim 3\%$ larger than those obtained with {\sc mepsa}. The 90 \% confidence interval is [0.4-2.3], meaning that for most GRBs, the discrepancy between the MVTs obtained using BBs and those obtained using {\sc mepsa} is smaller than a factor of 2.
We further estimated that in roughly $60\%$ cases, the temporal structures identified by {\sc mepsa} and BBs coincide; in such cases the discrepancies in MVT values arise only from different ways of estimating the width. BBs, which approximate the pulse as a rectangle, tend to overestimate the width, whereas {\sc mepsa} uses a more realistic, though simplified, pulse shape.

\begin{figure}[h!]
    \centering
    \includegraphics[width=0.9\linewidth]{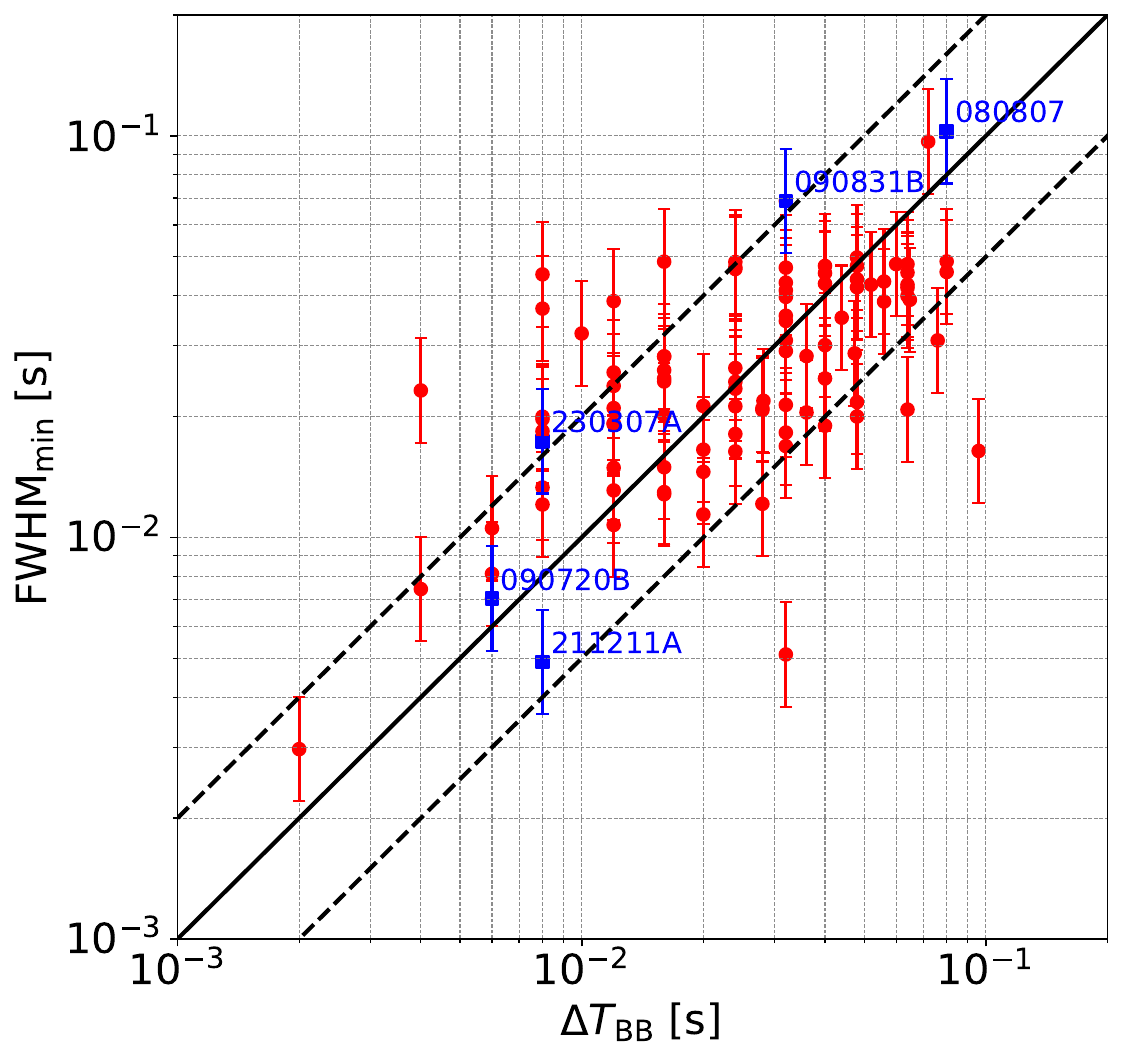}
    \caption{$\rm{FWHM}_{\rm{min}}$ vs the shortest segment of the BBs segmentation, ${\rm \Delta T_{BB}}$. Blue points represents the GRBs shown in Fig.~\ref{fig:see_grbs}.}
    \label{fig:mvt_vs_BBs}
\end{figure}
\begin{figure}[h!]
    \centering
    \includegraphics[width=0.8\linewidth]{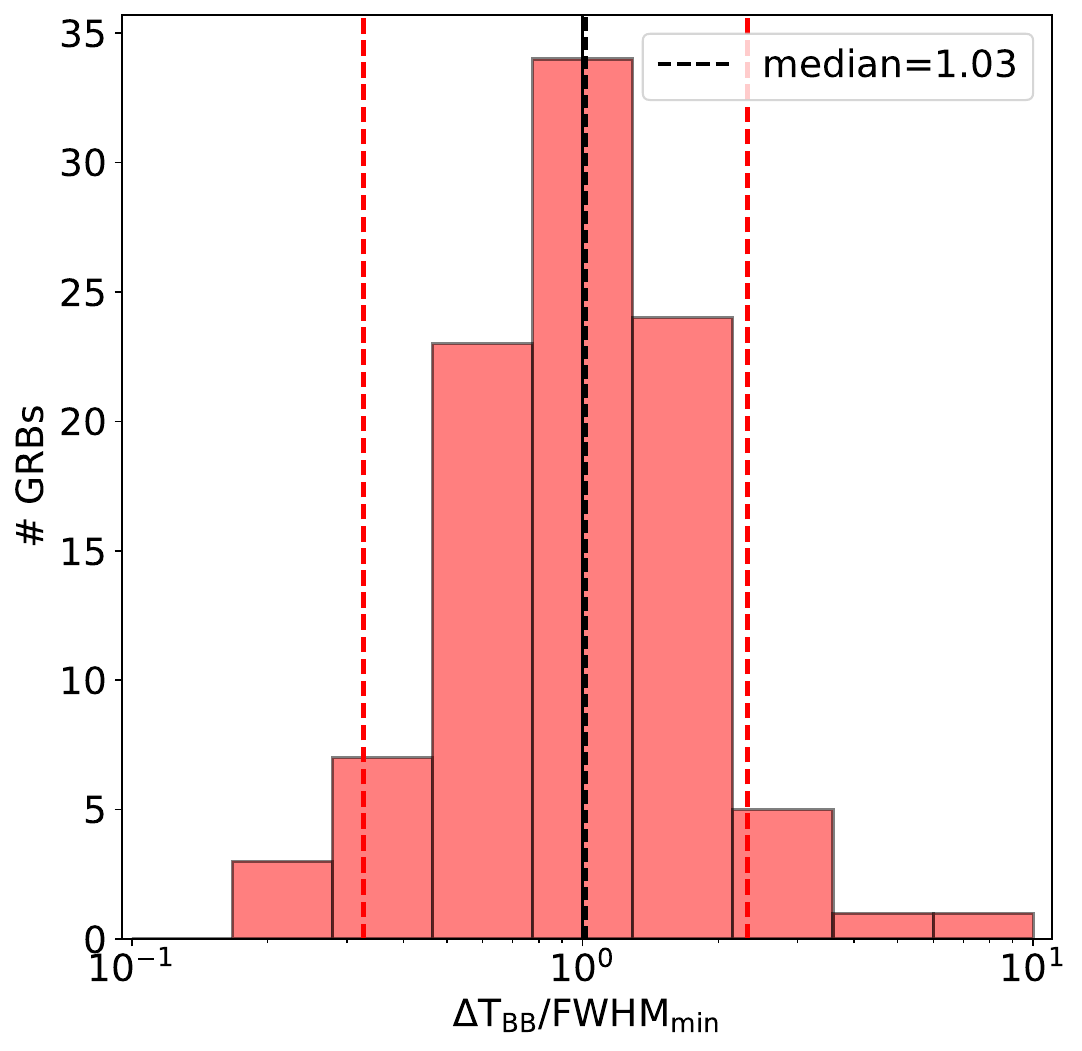}
    \caption{Distribution of the ratio ${\rm \Delta T_{BB}}$ over $\rm{FWHM}_{\rm{min}}$. The solid black line represents the case $\rm{FWHM}_{\rm{min}} = \Delta T_{BB}$, while the dashed one shows the median value. The two red dashed lines enclose the 90 $\%$ confidence interval [0.4-2.3].}
    \label{fig:mvts_vs_bbs_ratio}
\end{figure}
\FloatBarrier
\begingroup
\let\clearpage\relax
\let\cleardoublepage\relax

\section{Dependence of $\rm{FWHM}_{\rm{min}}$ on energy}
\endgroup
\label{sec:appendix_c}

We computed $\rm{FWHM_{\rm{min}}}$ as a function of the geometric mean of the energy range boundaries, for six different energy ranges: 8-30, 8-90, 30-90, 8-1000, 90-300, and 90-1000 keV.  We carried out this analysis on 286 bursts, each having a measured $\rm{FWHM_{\rm{min}}}$ with $\rm{S/N}>7$ across all six energy ranges. We found that $\rm{FWHM_{\rm{min}}} \propto E^{-\alpha}$ with $\alpha_{\rm{mean}} = 0.46 \pm 0.19$, with a standard dispersion of $\sigma=0.7$, and $\alpha_{\rm{median}} = 0.26\pm 0.12$. Results are shown in Figure ~\ref{fig:FWHM_vs_erange}. The results are consistent with those of \citetalias{Camisasca23} that obtained $\alpha_{\rm{mean}} = 0.45 \pm 0.08$, $\alpha_{\rm{median}} = 0.54\pm 0.07$ and those of \cite{Fenimore95}: $\alpha\in[0.37;0.46]$. 

\begin{figure}[h!]
    \centering
    \includegraphics[scale=0.25]{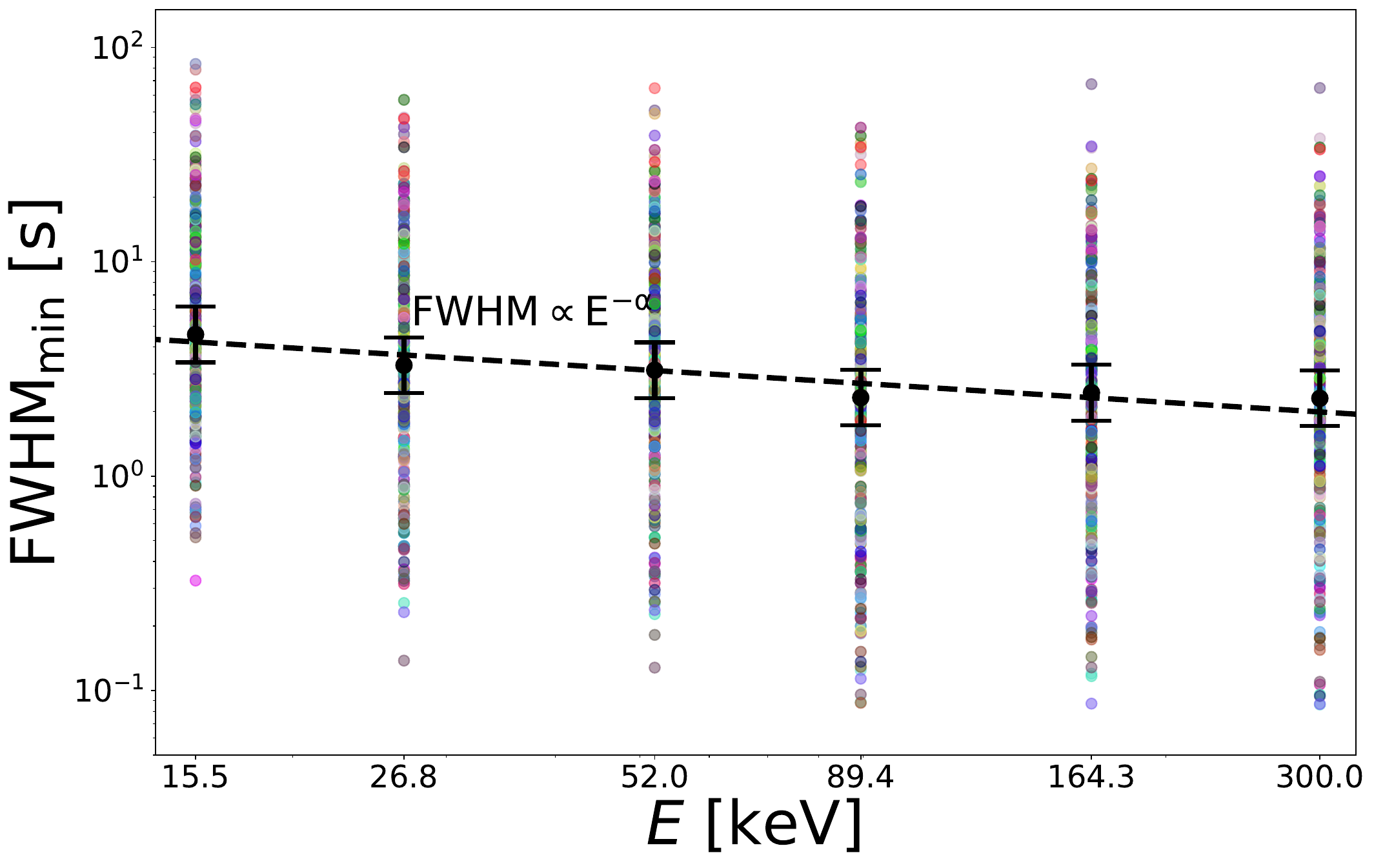}
    \caption{The coloured dots are the $\rm{FWHM_{\rm{min}}}$ of 286 bursts in four different energy ranges: 8-30, 8-90, 30-90, 8-1000, 90-300, and 90-1000 keV. The values on the $x$-axis are the geometric means of the corresponding energy boundaries. Black dots with error bars are the weighted averages of the $\rm{FWHM_{\rm{min}}}$ for each energy range and the black dashed line is the power-law that best fit the black points.}
    \label{fig:FWHM_vs_erange}
\end{figure}

\end{appendix}

\FloatBarrier
\let\cleardoublepage\clearpage
\end{document}